\documentclass[12pt,a4paper]{article}
\usepackage{amsmath, amssymb, bm}
\usepackage{booktabs}
\usepackage{geometry}
\usepackage{graphicx}
\usepackage{cite}
\usepackage{caption}
\usepackage{braket}
\usepackage[colorlinks=true, linkcolor=blue, citecolor=blue, urlcolor=blue]{hyperref}
\usepackage{titlesec}
\usepackage{indentfirst}
\usepackage{lineno}

\titleformat{\section}
{\normalfont\bfseries\large}
{\thesection.}{0.5em}{}

\usepackage{authblk}   
\usepackage{orcidlink}

\title{Electrically tunable spin qubits in strain-engineered graphene $p$-$n$ junctions}

\author[1,2]{Myung-Chul Jung\orcidlink{0000-0003-4133-7710}
\thanks{first author: \texttt{mcjung24@chosun.ac.kr}}}
\author[1,2]{Nojoon Myoung\orcidlink{0000-0003-2344-5793}\thanks{Corresponding author: \texttt{nmyoung@chosun.ac.kr}}}

\affil[1]{Department of Physics Education, Chosun University, Gwangju 61452, Republic of Korea}
\affil[2]{Institute of Well-Aging Medicare \& Chosun University G-LAMP Project Group, Chosun University, Gwangju 61452, Republic of Korea}

\date{}

\vspace{10pt}
\begin{document}
\maketitle
\newpage

\begin{abstract}
Strain engineering enables quantum confinement in pristine graphene without degrading its intrinsic mobility and spin coherence. Here, we extend previously proposed strain-induced charge-qubit architectures by incorporating spin degrees of freedom through Rashba spin-orbit coupling and Zeeman fields, enabling spin-qubit operation in single-layer graphene. In a graphene $p$-$n$ junction, a strain-induced nanobubble generates a pseudomagnetic field that forms double quantum dots with gate-tunable level hybridization. Tight-binding quantum transport simulations and a four-band model reveal two distinct avoided crossings: spin-conserving gaps at zero detuning and spin-flip gaps at finite detuning, the latter increasing with spin-orbit coupling strength while the former decreases. Time-domain simulations confirm detuning-dependent Rabi oscillations corresponding to these two operational regimes. These results demonstrate that strain-induced confinement combined with tunable spin-orbit coupling provides a viable mechanism for coherent spin manipulation in pristine graphene, positioning strained single-layer graphene as a promising platform for scalable spin-based quantum technologies.
\end{abstract}



\section{Introduction}
Graphene exhibits exceptional physical properties such as ultra-high carrier mobility, long relaxation times, and long spin coherence~\cite{Trauzettel2007,Oh2010,Han2012,Volk2013,Banszerus2022,Gachter2022,Garreis2024}. These characteristics make graphene an attractive platform for quantum information technologies. However, because single-layer graphene (SLG) lacks an intrinsic band gap, electrostatic carrier confinement is challenging, making it difficult to realize well-defined qubits.
To overcome this limitation, bilayer graphene (BLG) has been widely explored, as a tunable band gap can be opened by applying a perpendicular electric field. This property has enabled the realization of graphene-based quantum dots (QDs) and spin qubits~\cite{Zhang2009,Banszerus2022,Denisov2025,Duprez2024,Garreis2024}. Despite these advances, BLG generally exhibits reduced carrier mobility and weaker spin coherence compared to SLG. Therefore, achieving controllable confinement in SLG--while preserving its superior intrinsic electronic and spin properties--remains an important challenge.

A particularly promising route is strain engineering, which enables electronic confinement through mechanical deformation rather than electrostatic band-gap opening. Nonuniform strain in graphene induces pseudomagnetic fields (PMFs) that act as effective magnetic fields on Dirac fermions, leading to the formation of discrete Landau levels~\cite{Levy2010,Low2010,Guinea2010,Kitt2012,Zhu2015,Li2020,Guinea2010sci,Vozmediano2010,Nikolai2012,Zhu2014,Myoung2020}. This approach, often referred to as straintronics, allows one to tailor the electronic structure of graphene without introducing impurities or breaking lattice symmetry~\cite{Si2016,Pereira2009,Guinea2012,Sahalianov2019,Myoung2014,Myoung2020,Myoung2022,Myoung2024,Jun2025}. Experimentally, strain-induced PMFs have been observed in graphene nanobubbles, where Landau-level quantization emerges due to strong local deformation~\cite{Levy2010}. Subsequent advances have demonstrated that nanobubbles and nanowrinkles can host localized electronic states, forming strain-induced QDs or waveguides in SLG~\cite{Low2010,Guinea2010,Levy2010,Kitt2012,Zhu2015,Li2020,Vozmediano2010,Nikolai2012,Zhu2014,Lavor2020,Milovanovic2017,Qi2013,Volk2013}.

Building on these developments, strain-induced graphene double quantum dots (DQDs) have been proposed as a promising candidate for charge qubit~\cite{Park2023,Myoung2020}. To further advance this platform, in this study, we incorporate the spin degree of freedom (DOF) in the charge-based graphene qubit system, extending concepts that have been successfully demonstrated in III–V semiconductors and BLG systems~\cite{Zajac2018,He2019,Engel2004,Guttinger2010,Banszerus2022,Denisov2025}. In this context, SLG offers a distinct advantage due to its extremely weak intrinsic spin-orbit coupling (SOC) and hyperfine interaction, which are expected to yield long spin coherence times—an essential requirement for spin-qubit applications~\cite{Trauzettel2007,Recher2010,Konschuh2010,Kochan2017}.
Furthermore, the proposed platform provides multiple independent control parameters for spin manipulation. The Rashba SOC (RSOC) strength $\lambda_R$ can be tuned by applying a vertical electric field~\cite{Avsar2020,Dmitry2021}, while external magnetic fields allow independent control of the Zeeman splitting $\Delta_z$. Recently, proximity-induced effects in TMD heterostructures--such as SLG/WSe$_2$--offer a way to modulate the SOC strength in graphene~\cite{Avsar2014,Gmitra2015,Wang2016,Wang2015,Yang2017,Gerber2025,Masseroni2024,Kurzmann2021,Dulisch2025}. These complementary tuning knobs enable coherent spin manipulation without compromising graphene’s intrinsic electronic quality.

In this work, we establish a theoretical framework for realizing controllable spin qubits in strain-engineered SLG. Our model integrates pseudomagnetic confinement induced by local strain with tunable RSOC and Zeeman fields, enabling coherent manipulation of spin states in pristine SLG. We show that the qubit operation can be tuned through mechanical (bubble geometry), electrical (detuning), and magnetic (SOC and Zeeman) parameters, allowing mode-selective control within a unified device architecture. We further analyze the resulting energy spectra and Rabi dynamics to identify distinct operational regimes, laying the groundwork for strain-based spin qubits in scalable 2D quantum platforms.

This paper is organized as follows. Section~\ref{overview} introduces the theoretical model and provides an overview of the proposed spin-qubit system. Section~\ref{quantum_cond} presents the quantum conductance spectra, including spin-resolved effects. Section~\ref{detuning} analyzes the detuning spectra and the associated gap behavior. Section~\ref{hamil} introduces the effective four-band Hamiltonian and examines its symmetry properties. Section~\ref{Rabi osc} presents the Rabi oscillation map as a function of the RSOC strength. Section~\ref{discussion} discusses readout schemes and scalability. Finally, Section~\ref{conclusion} summarizes the main results.

\begin{figure}[htbp]
 	\centering
 	\includegraphics[width=0.8\columnwidth]{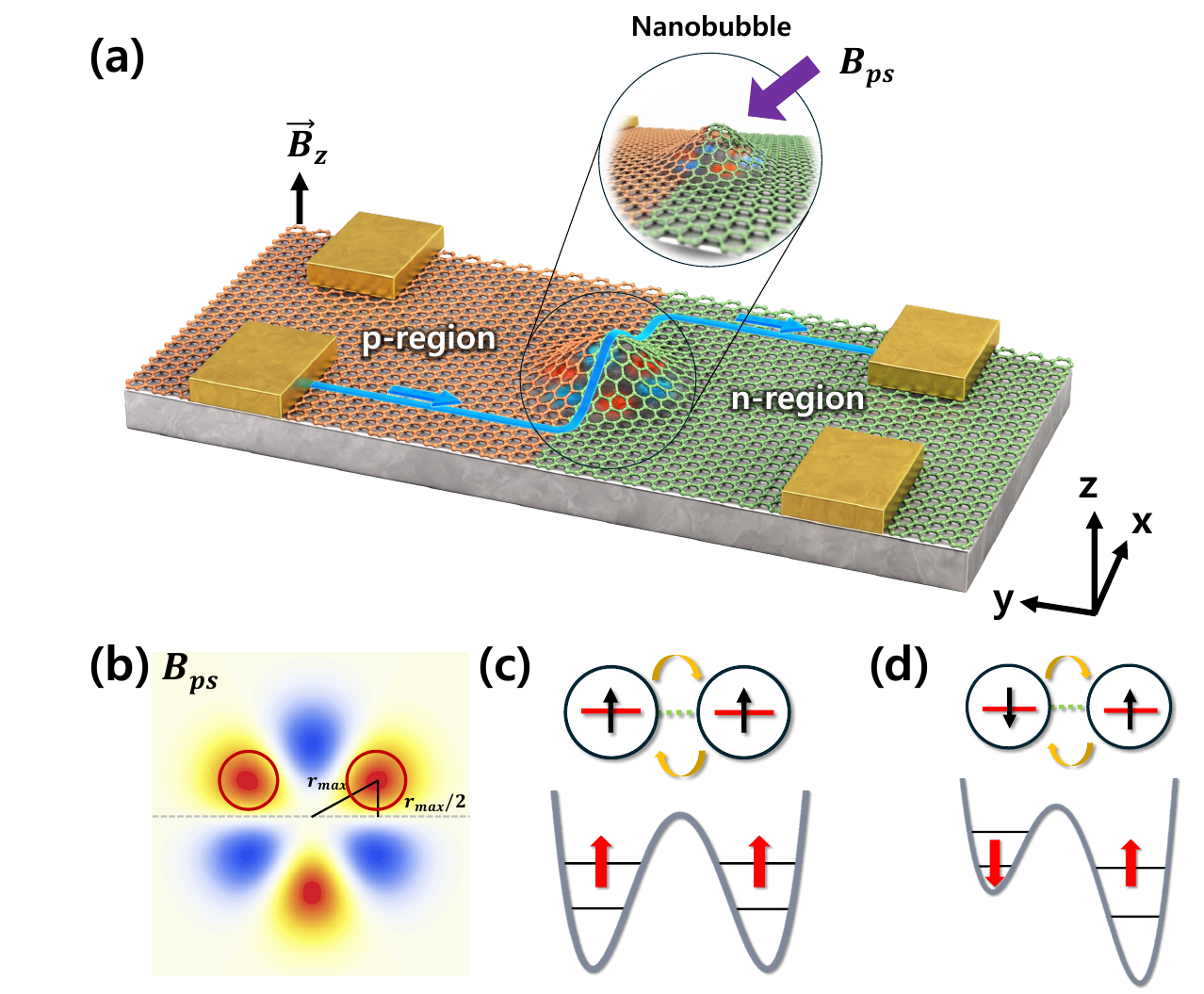}
 	\caption{\label{fig1} 
    Schematics of the strain-engineered graphene $p$-$n$ junction spin-qubit device and detuning-induced spin transitions. 
    (a) Device geometry of the strained graphene nanobubble located at the interface of the $p$-$n$ junction under a perpendicular external magnetic field $\vec{B}_{z}$. The localized strain creates a confined potential landscape within the nanobubble region. The blue line shows a quantum Hall current flow. 
    Inset: Enlarged view of the graphene nanobubble illustrating the locally induced strain and curvature. The strain-induced confinement generated by the associated pseudomagnetic field (PMF) is depicted beneath the nanobubble.
    (b) Profile of the PMF $B_{ps}$ in the region of the nanobubble. The dashed line marks the junction interface, and $r_{\mathrm{max}}$ represents the distance from the PMF center.
    (c) Symmetric double-quantum-dot potential and parallel spin configuration in the absence of electrical detuning (spin-conserving transition). 
    (d) Asymmetric confinement potential induced by detuning, enabling hybridized interdot tunneling accompanied by spin-flip transitions mediated by Rashba spin-orbit coupling.
    }
\end{figure}

\section{Results}
\subsection{Overview of strain-engineering graphene spin qubit}
\label{overview}
To investigate the formation and transport properties of spin qubits in strained graphene, we consider a theoretical model based on a $p$-$n$ junction with a strain-induced nanobubble (NB) under a homogeneous external magnetic field $B_{z}$ (see Supplementary Materials Section 10 for NB profile). As shown in Fig. \ref{fig1}(a), a uniform NB is formed at the $p$-$n$ junction interface, and this strain-induced gauge field leads to local confinement, creating a QD-like potential minimum within the NB as shown in Figure \ref{fig1}(b).
Note that we consider a split bottom-gate geometry to control the $p$-$n$ junction potential, as the effect of NB-induced electrostatic non-uniformity is smaller than the relevant energy scales used in our model.

The blue arrow in Fig.~\ref{fig1}(a) represents the charge transport across the $p$-$n$ junction in a perpendicular magnetic field, where carrier trajectories exhibit a path reversal at the interface (See Supplementary Materials Section 1 for more details). 
In the $n$-region, electrons follow cyclotron orbits with a given chirality, while in the $p$-region, holes circulate in the opposite direction. 
As a result, the carriers form snake-like trajectories along the $p$-$n$ junction, propagating along the interface due to the opposite curvature of their cyclotron motion~\cite{Milovanovic2014}.
These interface states play a crucial role in mediating the coupling between the propagating quantum Hall (QH) channel and the localized QD states induced by the PMF, as discussed previously~\cite{Myoung2020,Park2021}. The degree of hybridization between the QH channel and the QD is determined by their spatial overlap, which is governed by the magnetic length $l_B = \sqrt{(\hbar/eB)}$. When the distance between the QD localization region and the interface becomes comparable to $l_B$, significant hybridization occurs, leading to resonant transport features. Conversely, when this separation exceeds $l_B$, the coupling is suppressed, and the QD remains more localized.

Figure \ref{fig1}(c) shows the case of a symmetric potential profile due to no detuning. In this regime, the relevant energy levels retain the same spin orientation, leading to a spin-conserving avoided crossing (see Section \ref{detuning}). Such a configuration supports qubit operations that rely on transitions within the same spin manifold, effectively realizing a charge-qubit-type operation.
In contrast, Fig.~\ref{fig1}(d) illustrates the behavior under an asymmetric potential. The detuning introduced by the gate voltage modifies the electrostatic confinement on one side of the junction, carrying the opposite-spin energy levels into resonance. This condition results in a spin-flip avoided crossing, where interdot tunneling hybridizes quantum states with opposite spin orientations (see Section \ref{detuning}). The resulting spin-flip coupling enables electrically driven spin control mediated by spin-orbit interaction, realizing a spin-qubit-type operation.

Together, these results demonstrate that strain-induced confinement, combined with gate-controlled potential asymmetry, provides a versatile and unified platform for engineering spin-conserving (charge qubit) and spin-flip (spin qubit) operations in graphene. The interplay among PMFs, electrostatic gating, and spin-dependent hybridization establishes the physical foundation for the coherent spin manipulation investigated in the subsequent sections of this work.

\subsection{Quantum conductance}
\label{quantum_cond}
Now, to investigate how the inclusion of the spin DOF modifies the characteristics of the recently proposed strained graphene qubit system~\cite{Myoung2020,Park2023}, we first examine the quantum conductance map as a function of $h_{0}$, which scales directly with the PMF induced by the NB. 

\begin{figure}[htbp]
 	\centering
 	\includegraphics[width=\columnwidth]{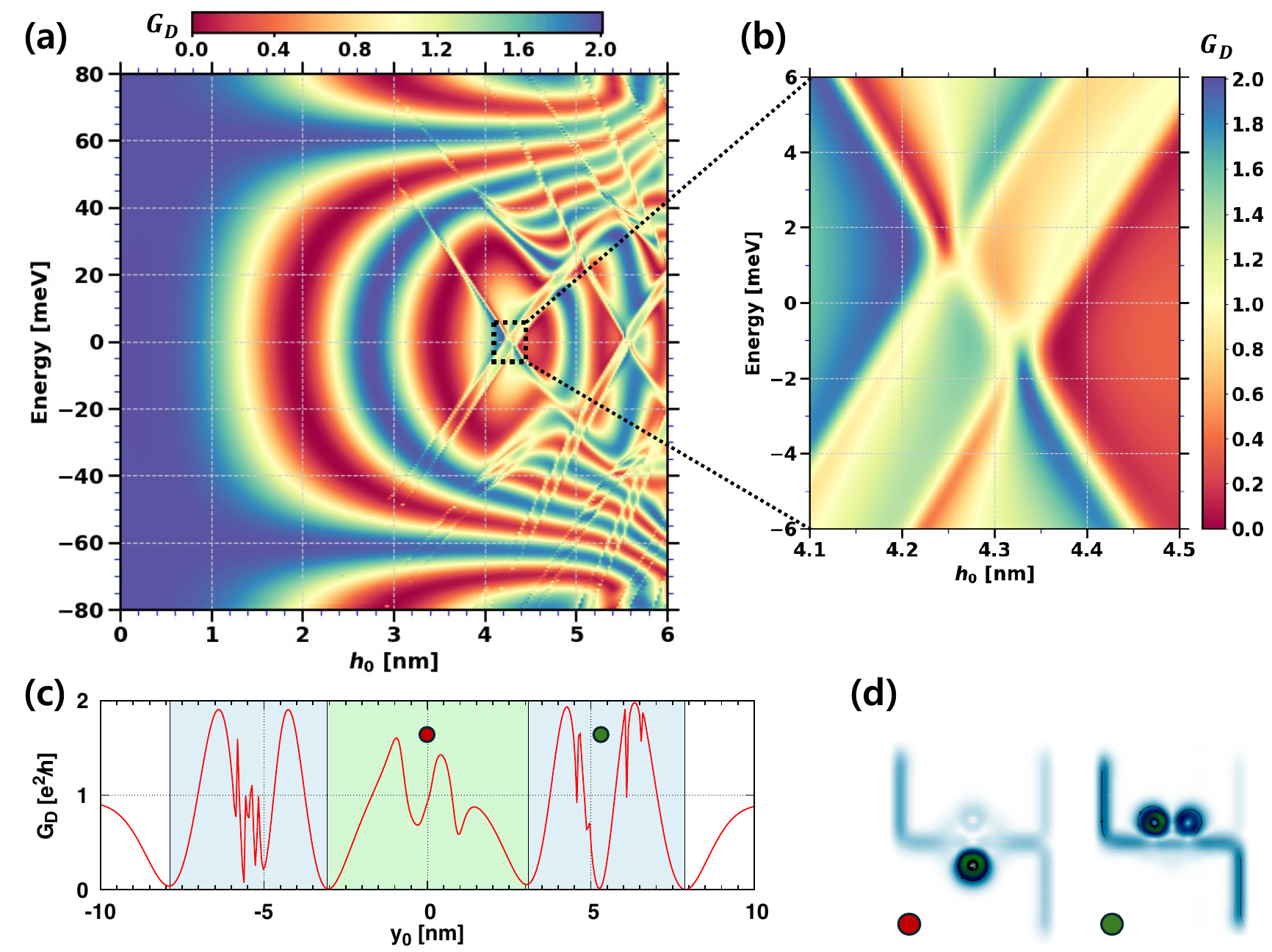}
 	\caption{\label{fig2} 
    Quantum conductance including RSOC and the Zeeman field. 
    (a) Quantum conductance map under the Zeeman field ($\Delta_z=1.7~meV$) with RSOC ($\lambda_R=10.1~meV$). 
    (b) The zoomed-in views around the single quantum dot ground level in the range of $h_{0}=4.1$ to $4.5$ nm. 
    (c) Quantum conductance line as a function of $y_0$, representing the position of the $p$-$n$ interface channel at $h_0=4.3~nm$. 
    (d) The current density plots corresponding to the colored circles in (c).
    }
\end{figure}

Figure~\ref{fig2}(a) shows the conductance resonances with RSOC in the quantum conductance $G_{D}$ characteristics of the strained graphene system. In the absence of RSOC, the Zeeman field ($\Delta_z=1.7~meV$) produces two distinct spin-split branches, leading to independent crossing Fano resonance lines (see Supplementary Figure S2).
When RSOC ($\lambda_R=10.1~meV$) is taken into account, the SOC mixes the spin states.
Therefore, the inclusion of RSOC modifies the conductance resonances in the strained graphene qubit system, as shown in Figure~\ref{fig2}(a). 
In Fig.~\ref{fig2}(b), the zoomed-in regions around the single quantum dot (SQD) regime are highlighted in Fig.~\ref{fig2}(a) for the case with RSOC, which clearly exhibit the gap evolution near the ground-state level.
As shown in Fig.~\ref{fig2}(b), the resonance lines become an asymmetric structure in the presence of RSOC, indicating the Rashba spin-orbit-induced modification of the interference pattern, while the resonance lines remain symmetric when only the Zeeman field is present, without RSOC (see Supplementary Figure S2(a)). 
Therefore, RSOC not only alters the Zeeman-induced spin splitting but also can be a tunable parameter to control quantum interference in strained graphene-based qubit devices.

We note that the external magnetic field employed in this study is relatively large ($B_{\mathrm{z}} \sim 30$~T), as required to access the quantum Hall regime within a finite-size system. This choice arises from the need to match the magnetic length to the confinement scale of the nanostructure, enabling strong coupling between the interface channel and the PMF-induced localized states while maintaining sufficiently large Landau-level separation in the quantum Hall regime. Importantly, the underlying qubit physics is governed not by the absolute magnitude of the magnetic field, but by the ratio between the magnetic length and the confinement scale. Consequently, for larger device sizes, the same physical regime can be achieved with proportionally reduced magnetic fields ($\sim 7$~T). To verify this, we adopt a recently proposed scaling approach~\cite{Liu2015,Liu2026} and confirm that equivalent results are reproduced at lower magnetic fields in larger systems (see Supplementary Materials Section 3 for details). These results indicate that the predicted qubit behavior is not intrinsically tied to high magnetic fields, but can be realized under experimentally realistic conditions in appropriately scaled strained-graphene devices.

Figure \ref{fig2}(c) displays the line cuts of the quantum conductance $G_{D}$ as a function of the interface channel position $y_{0}$ for $h_0$=4.3 nm and $\sigma$=7.38 nm, including both RSOC and Zeeman coupling.
Clearly, $G_D$ with RSOC exhibits an asymmetric shape, as shown in Fig.~\ref{fig2}(a). The green area is the SQD activation region, and the blue area is the DQD activation region, respectively.
The colored symbols in Fig.~\ref{fig2}(c) correspond to selected points where the associated current density patterns are shown in Fig.~\ref{fig2}(d), illustrating the spatial localization and transport channels of the single and double quantum dot states.

The PMF has opposite signs in the $K$ and $K'$ valleys (see Supplementary Fig.~S8(b)). However, the intervalley scattering in our system is absent, as established in previous studies on strain-induced gauge fields in graphene~\cite{Myoung2020}. 
Therefore, the two valleys remain decoupled and can be treated as independent subspaces. Furthermore, as shown in Supplementary Fig.~S8(a), the valley splitting is much larger than the relevant energy scales in our model, including the RSOC and the Zeeman splitting. 
Consequently, because intervalley coupling can be safely neglected, and a single-valley description is well justified, we focus on the qubit properties defined within the single $K$ valley.

\subsection{Detuning energy spectrum}
\label{detuning}

\begin{figure}[htbp]
 	\centering
 	\includegraphics[width=\columnwidth]{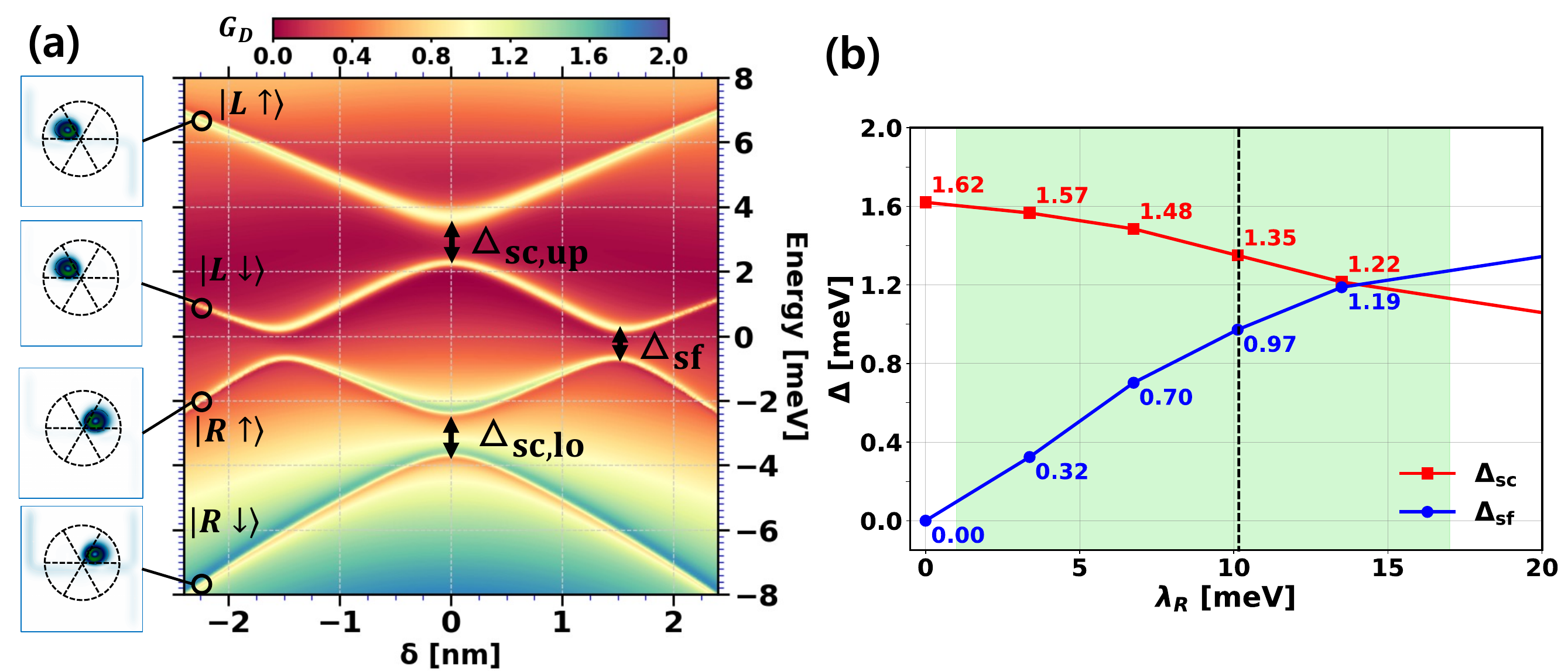}
 	\caption{\label{fig3} 
    Detuning-dependent energy spectrum and gap evolution for $h_0$=4.29 nm, $\sigma$=7.38 nm, and $y_0$=5.2 nm.
    (a) Detuning energy spectrum of the double quantum dot (DQD) as a function of the detuning parameter $\delta$, evaluated at $\lambda_R = 10.1~\mathrm{meV}$ and $\Delta_z = 1.7~\mathrm{meV}$, corresponding to the solid line in Fig.~3(b).
    The labels $L$ and $R$ denote the left and right QD states at $K$-valley, respectively, and the $\uparrow$ and $\downarrow$ indicate the spin states.
    The gaps $\Delta_{\mathrm{sc}}$ and $\Delta_{\mathrm{sf}}$ correspond to the energy splittings at zero detuning ($\delta=0$) and at finite detuning ($\delta$ = $\delta_0$), respectively. For $\Delta_{\mathrm{sc}}$, $up$ and $lo$ denote the upper and the lower position of energy, respectively.
    The left blue panels illustrate the current density distribution for each eigenstate.
    (b) Evolution of the energy gaps $\Delta$ as a function of $\lambda_R$ , obtained from numerical simulations. Here, $\Delta_{\mathrm{sc}}$ is the average value of $\Delta_{\mathrm{sc,up}}$ and $\Delta_{\mathrm{sc,lo}}$. The green area corresponds to the value of RSOC strength (1--17 $meV$) measured in the SLG/TMD heterostructure\cite{Avsar2014,Gmitra2015,Wang2016,Wang2015,Yang2017,Gerber2025,Masseroni2024,Kurzmann2021,Dulisch2025}.
    }
\end{figure}

To understand the operation of the strained graphene spin qubit, we analyze the detuning effect in a DQD configuration (located at $y_0$=5.2 nm). The detuning is introduced by modifying the electrostatic potential to represent an asymmetric $p$-$n$ junction~\cite{Williams2007,Beenakker2007,Wang2012,Kochan2017,Pereira2009}, 
\begin{equation}
\label{pn_profile}
u_i = u_0 \tanh\!\left[\frac{y_i + 0.5\delta \,(1+\tanh x_i)}{\xi}\right],
\end{equation}
where $\delta$ denotes the displacement of the $p$-$n$ junction interface, $\xi\sim30a_0$ sets the characteristic width (i.e., sharpness) of the interface, and $u_0=2\sqrt{2e \hbar v_F^2 B}$ represents the energy separation between the zeroth and first Landau levels in graphene under a homogeneous external magnetic field $B_{z}$. 
This potential profile captures the realistic effect of charge screening in experimental devices. Within this framework, the junction width $\xi$ serves as a characteristic length scale governing the spatial variation of the potential, and our analysis suggests that a well-defined DQD potential can be maintained over a finite range of $\xi$. 
As discussed in previous studies~\cite{Myoung2020,Park2021}, the junction width controls the coupling between the quantum Hall (QH) interface channel and the localized DQD states: for a sharper junction (small $\xi$), the interface channel approaches the dot region, leading to stronger hybridization and broader Fano resonances, whereas for a smoother junction (large $\xi$), the spatial separation increases, resulting in reduced coupling and sharper resonances. 
This behavior can be understood from the relative positioning between the QD minima and the QH interface channel. The minimum points of QD are formed near the region where the magnitude of PMF is maximized, which occurs at a finite radius $r_{\mathrm{max}} \sim \sqrt{3/2}\,\sigma$ from the center of the NB (See Figure~\ref{fig1}(b)). Therefore, the effective coupling is governed by the relative separation between this PMF-induced localization region and the interface channel. When $\xi$ becomes comparable to $r_{\mathrm{max}}$/2, the coupling between the DQD and the channel smoothly increases.
Accordingly, we choose $\xi$ within a regime where the resonant features are clearly resolved, ensuring stable and well-defined QD formation while maintaining finite coupling to the interface channel.

As detuning $\delta$ increases, the two QDs become inequivalent, as shown in Fig.~\ref{fig1}(d), leading to the breakdown of the symmetric superposition states.
In the energy spectrum of Fig.~\ref{fig3}(a), two distinct sets of resonance lines with different splitting appear, similar to the conductance maps discussed earlier in Section \ref{quantum_cond}.
Because the spin-down states, $|$$\downarrow$$\rangle$, lie at lower energy than the spin-up state, $|$$\uparrow$$\rangle$, we identify the four levels as $|L\uparrow\rangle$, $|L\downarrow\rangle$, $|R\uparrow\rangle$, and $|R\downarrow\rangle$ as shown in Fig.~\ref{fig3}(a). The labels $L$ and $R$ are assigned by analyzing the current density associated with each resonance line. Two types of well-defined avoided crossings emerge, giving rise to two characteristic detuning-induced gaps: the spin-conserving gap ($\Delta_{\mathrm{sc}}$) and the spin-flip gap ($\Delta_{\mathrm{sf}}$).
For the zero detuning gaps, $\Delta_{\mathrm{sc,up}}$ and $\Delta_{\mathrm{sc,lo}}$, the avoided crossings involve the same spin states ($|L\uparrow\rangle$ with $|R\uparrow\rangle$, or $|L\downarrow\rangle$ with $|R\downarrow\rangle$). In contrast, $\Delta_{\mathrm{sf}}$ is defined by the minimum energy separation between the middle bands, corresponding to avoided crossings between different spin states.
For the operation of the spin qubit, it is crucial to understand how the detuning-induced gaps evolve under the influence of RSOC. Under the fixed Zeeman energy $\Delta_z$, we evaluated $\Delta_{\mathrm{sc}}$ and $\Delta_{\mathrm{sf}}$ as functions of $\lambda_R$ as shown in Fig. \ref{fig3}(b). 
$\Delta_{\mathrm{sc}}$ gradually decreases with increasing $\lambda_{R}$, although RSOC does not explicitly contribute to its gap behavior. Note that this gap originates from avoided crossings between same-spin states. 
To physically understand this origin, we analyzed the characteristics using the effective potential, written as 
\begin{equation}
\label{eff_pot}
V_{eff} = v^2_{F}e\hbar[B_{ps,z}(r)\tau_0\otimes\sigma_z - B_z\tau_z\otimes\sigma_z] + U_0\tanh[y/\xi],
\end{equation}
where $\tau$ denotes the Pauli matrices for the isospin and $\tau_{0}$ is the $2 \times 2$ identity matrix. 
Here, $B_{ps}$ denotes the strain-induced pseudomagnetic field generated by the graphene nanobubble, whereas $B_z$ refers to the homogeneous externally applied magnetic field.
As shown in Supplementary Figure S4(a), the energy at which the avoided crossing of $\Delta_{\mathrm{sc}}$ appears shifts to higher values with increasing RSOC strength. This behavior can be interpreted as an effective shift of the $p$-$n$ junction interface toward the DQD region, arising from RSOC-induced modifications of the effective potential landscape (see Supplementary Materials Section 6 for details). As the interface approaches the DQD, the effective confinement potential is altered, leading to an increased spatial separation between the two local minima defining the QDs. This enhanced separation reduces the interdot tunnel coupling as shown in Supplementary Figure S6(c). Since the observed gap at $\Delta_{\mathrm{sc}}$ originates from interdot hybridization rather than a direct RSOC-induced band gap, the reduction in interdot coupling results in a smaller avoided crossing gap. Therefore, although RSOC generally introduces spin-dependent energy scales, in this system, its net effect is to weaken the hybridization-induced gap by effectively increasing the interdot separation.
In contrast, $\Delta_{\mathrm{sf}}$ arises from avoided crossings between opposite-spin states, where RSOC plays a dominant role in inducing hybridization. Consequently, as shown in Fig. \ref{fig3}(b), $\Delta_{\mathrm{sf}}$ increases with $\lambda_{R}$, while $\Delta_{\mathrm{sc}}$ decreases. Additionally, the detuning position of the $\Delta_{\mathrm{sf}}$ shifts toward larger detuning values as the RSOC strength increases (see Supplementary Figure S4).

Here, we focus on the sweet spots in the detuning energy spectrum and identify two distinct operating regimes characterized by different noise sensitivities. 
The issue of noise sensitivity at sweet spots has been extensively discussed in the context of conventional flopping-mode qubits, where achieving stable operation requires minimizing sensitivity to charge fluctuations~\cite{Petersson2012,Teske2023,Ungerer2024,Hajati2024,Young2025,Noirot2025,Losert2025,Stastny2025,Chan2018,Yoneda2018,Benito2017,Benito2019a,Benito2019b,Cayao2020}. 
Therefore, operating at noise-resilient sweet spots is essential for reliable qubit manipulation. 

To assess this in our system, we analyze the sensitivity to charge noise. 
Although two avoided crossings in our system are relatively narrow and may appear more susceptible to charge noise, our analysis shows that the resulting fluctuation scale remains small under realistic conditions. 
To quantify this, we estimate the effect of charge noise by considering a typical fluctuation amplitude of $\delta u_0 \sim 10~\mu$eV. Using the relation $\delta y = (\xi/u_0)\delta u_0$, we obtain an effective fluctuation scale of approximately $\sim 1~\mathrm{pm}$ around both the zero-detuning and finite-detuning operating points. 

These results demonstrate that the system remains robust against realistic charge noise. 
In this regard, unlike conventional approaches that operate in the tunneling-dominated regime and rely on externally driven fields to induce spin-charge hybridization~\cite{Teske2023,Benito2019a,Benito2019b,Cayao2020,Losert2025,Benito2017}, our scheme exploits intrinsic RSOC to enable spin manipulation through detuning control alone, as previously proposed in Refs.\cite{Bulaev2007,Mutter2021,Hajati2024}.
Rather than replacing conventional flopping-mode strategies, our proposal provides an alternative operating regime in strain-engineered graphene, offering a simplified platform for spin manipulation while maintaining reasonable robustness against charge noise.
\\

\subsection{Effective Hamiltonian for DQDs}
\label{hamil}
To describe the low-energy detuning spectrum and the associated avoided crossings, we construct a reduced four-band Hamiltonian projected onto the localized basis states \(|L,\uparrow\rangle, |L,\downarrow\rangle, |R,\uparrow\rangle, |R,\downarrow\rangle\), which capture the essential low-energy physics of the strained graphene spin-qubit system.
To analyze the gap behavior, the Hamiltonian is given by
\begin{equation}
\label{equation3}
    H(\delta,\Delta_z,\lambda_R,t_c) = \frac{\alpha\delta}{2}\tau_z+ t_c\tau_x + \frac{\Delta_z + \gamma\lambda_R^2}{2}\sigma_z + \frac{\beta\lambda_R}{2}\tau_y\sigma_y ,
\end{equation}
where $\delta$ is the detuning parameter, $t_c$ is the interdot tunnel coupling, $\Delta_z$ denotes the Zeeman splitting due to $B_z$, and $\lambda_R$ is the RSOC strength. Here, $\tau_i$ denote the Pauli matrices acting on the double-dot subspace, while $\sigma_i$ represent the Pauli matrices acting on the real-spin DOF. The coefficients $\alpha$, $\beta$, and $\gamma$ are fitting parameters determined by the best fit of the tight-binding four-band spectrum shown in Supplementary Fig.~S4(a), characterizing the strength of detuning, the Rashba interaction, and effective Zeeman splitting, respectively.
Each term captures a distinct physical mechanism in the DQD system, reflecting the same underlying physics as that previously studied in flopping-mode qubit systems~\cite{Yu2023,Mutter2021,Benito2019}.
The first term describes the detuning between the left and right quantum dots, controlling the energy offset between the localized states $|L\rangle$ and $|R\rangle$. 
The second term represents the interdot tunnel coupling, which hybridizes the left and right dot states and enables charge delocalization across the DQD. 
The third term corresponds to the Zeeman splitting, including an additional effective contribution arising from RSOC-induced second-order correction (See Supplementary Materials Section 8 for more details). 
The last term describes the RSOC, which couples the orbital (dot) and spin degrees of freedom and enables spin-flip processes accompanied by interdot transitions, playing a central role in coherent spin manipulation.
The interplay between detuning, tunnel coupling, Zeeman splitting, and RSOC gives rise to two distinct operational regimes: a spin-conserving regime near zero detuning and a spin-flip regime at finite detuning, providing multiple controllable pathways for qubit operation.

We emphasize that this Hamiltonian describes only a restricted subspace of the full system. Therefore, the projected Hamiltonian is not expected to preserve the full time-reversal symmetry of the original graphene system by itself, even in the absence of the external magnetic field ($\Delta_z=0$). In contrast, the Hamiltonian of the full system should preserve TRS (See Supplementary Materials Section 7 for more details).

As stated earlier, we extend a previously established charge-qubit platform~\cite{Myoung2020,Park2023} by incorporating the spin DOF in graphene.
The key ingredients for qubit manipulation are the Zeeman interaction and RSOC, whose physical origin and tunability are discussed below.
In pristine graphene, the intrinsic RSOC strength is on the order of a few $\mu$eV and is therefore negligibly small~\cite{Min2006}. However, recent studies have demonstrated that the SOC--Ising-type SOC and Rashba-type SOC--can be significantly enhanced via proximity effects in graphene–TMD heterostructures, reaching the meV scale~\cite{Avsar2014,Gmitra2015,Wang2016,Wang2015,Yang2017,Gerber2025,Masseroni2024,Kurzmann2021,Dulisch2025}. 
These heterostructures provide a practical and tunable route for controlling the spin-orbit interaction and the associated spin-flip gap $\Delta_{\mathrm{sf}}$. 
We note that proximity-induced SOC in graphene/TMD heterostructures generally consists of both Rashba and Ising-type contributions. 
The Ising SOC can be described as a valley-dependent effective Zeeman field of the form $\lambda_I \tau_z \sigma_z$, which induces spin splitting within each valley while preserving TRS.
Within a single valley, this term effectively acts as a Zeeman-like energy shift. Therefore, in our effective model, its contribution is incorporated into the Zeeman term $\Delta_z$, allowing us to focus explicitly on RSOC as the primary mechanism responsible for spin mixing and qubit control.
Although the relative strengths of Rashba and Ising SOC depend on the specific TMD material and interface conditions, the essential qubit operation mechanism relies on the presence of a finite RSOC that enables spin mixing. 
While this requirement is, in principle, not limited to a specific material, in practice only certain TMDs--most notably WSe$_2$--are known to provide a sufficiently strong Rashba component~\cite{Wang2015,Denis2017}. Therefore, its experimental realization requires TMD substrates that can induce a sizable RSOC.

Importantly, we emphasize that all four parameters in the Hamiltonian are independently tunable in experiments, providing multiple control knobs for manipulating the qubit system.
The detuning $\delta$ can be controlled by gate voltages, while Zeeman splitting $\Delta_z$ is determined by the external B-field.
In addition, the geometry of the graphene NB provides an additional tuning parameter for the DQD coupling $t_c$. In particular, elongation along the $x$-direction ($\sigma_x$) redistributes the PMF and modifies the confinement potential, effectively changing the separation between the two QDs~\cite{deformNB}. As a result, the interdot tunnel coupling is modulated, which is reflected in the variation of the avoided crossing in the detuning spectra~\cite{deformNB}.
\\

\subsection{Rabi oscillation}
Rabi oscillations provide a direct measure of the coherent controllability of a qubit. By computing the Rabi dynamics of the strained-graphene DQD, we verify whether the spin-conserving and spin-flip avoided crossings support coherent qubit rotations and how their operation frequencies evolve with RSOC, detuning, and Zeeman coupling using the Lindblad master equation. This analysis establishes the conditions under which the proposed system can function as a practical spin qubit.

Solving the Rabi dynamics requires a description of the time evolution of the driven two-level system under decoherence. For this purpose, we employ the Lindblad master equation, which captures both coherent evolution and dissipative relaxation processes. The master equation is given as
\begin{equation}
\frac{d\rho}{dt}
= -\frac{i}{\hbar}\,[H(t),\rho]
+ \sum_{k}\left( L_k \rho L_k^{\dagger}
- \frac{1}{2}\{ L_k^{\dagger} L_k , \rho \} \right),
\end{equation}
where $\rho(t)$ is the reduced density matrix of the two-level subspace, $H(t)$ is the time-dependent Hamiltonian extracted from the four-band model, and $L_k$ represents relaxation and dephasing channels included to model environment-induced decoherence.
Solving this equation yields the driven spin dynamics from which the Rabi oscillation frequency is extracted.

\label{Rabi osc}
\begin{figure}[htbp]
 	\centering
 	\includegraphics[width=1.0\columnwidth]{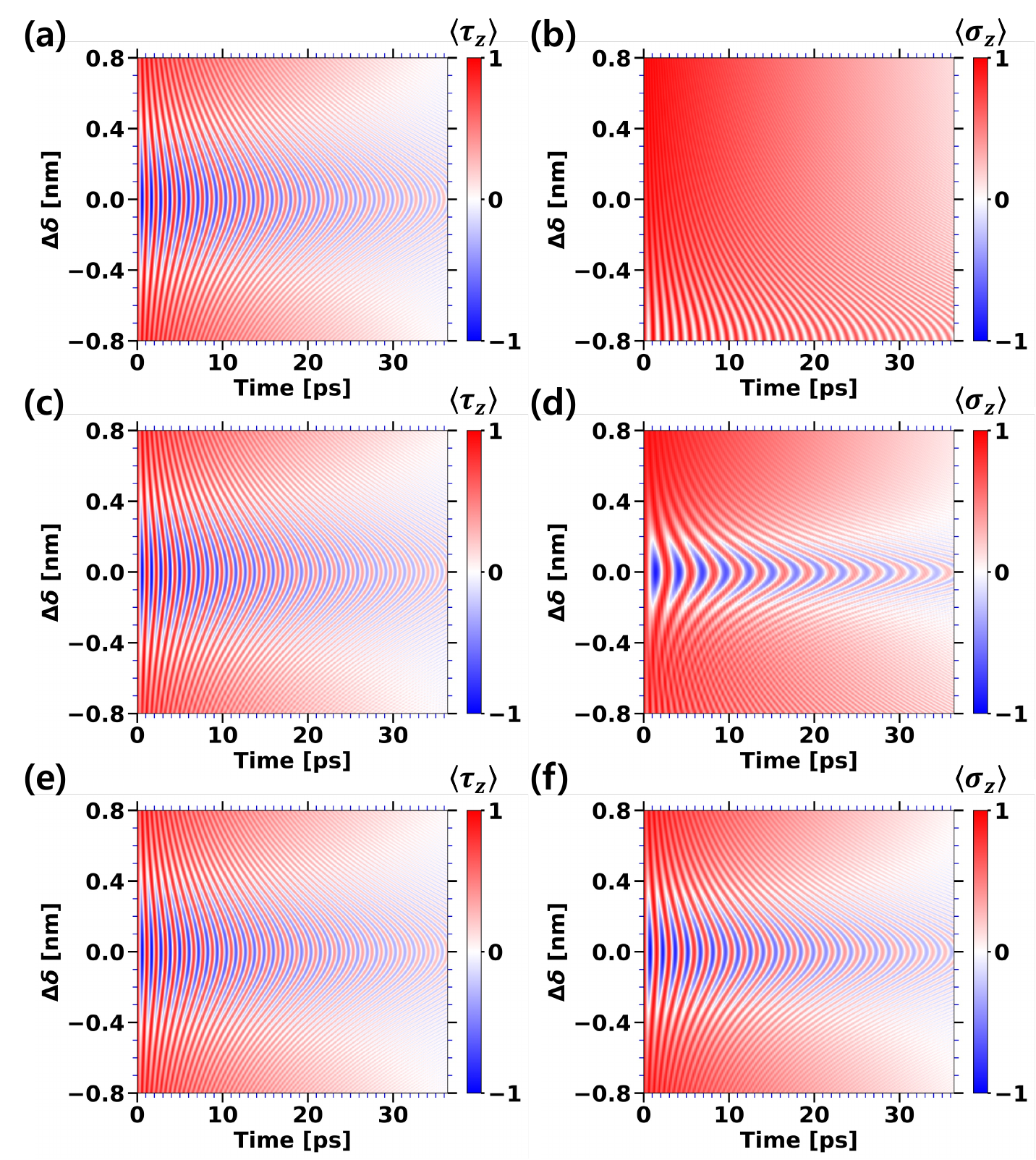}
 	\caption{\label{fig4} 
    Rabi oscillation maps as a function of RSOC strength $\lambda_R$.
    Rabi oscillation maps are shown for the spin-conserving regime $\delta$ = 0 (a,c,e) and the spin-flip regime $\delta$=$\delta_0$ (b,d,f). From top to bottom, the strength of $\lambda_R$ corresponds to 0, 6.75, and 13.5 meV, respectively. Each color map displays the Rabi frequency as a function of the driving detuning energy. 
    The horizontal axis represents the evolution time, and the vertical axis denotes the detuning offset $\Delta \delta$.}
\end{figure}

Figure \ref{fig4} presents the calculated Rabi oscillation maps for both the spin-conserving ($\delta$ = 0) and spin-flip ($\delta$ = $\delta_0$) operating regimes. Each map shows the time evolution of the spin polarization $\langle \tau_z(t) \rangle$ for spin-conserving and $\langle \sigma_z(t) \rangle$ for spin-flip as a function of the detuning parameter $\Delta \delta$. 
Here $\langle \tau_z(t) \rangle$ represents the isospin DOF describing the charge occupation between the two quantum dots. In contrast, $\langle \sigma_z(t) \rangle$ denotes the z-component of the real electron spin. When the SOC is turned off, the real spin remains conserved and does not undergo transitions. Therefore, $\langle \sigma_z(t) \rangle$ remains constant, while the observed oscillation originates from coherent charge dynamics between the two dots, described by the isospin expectation value $\langle \tau_z(t) \rangle$.
For the spin-conserving regime (See Figs.~\ref{fig4}(a), (c), and (e)), the oscillation frequency is maximal near $\Delta \delta$ $\sim$ 0, where the interdot tunneling is strongest, and monotonically decreases as detuning $\delta$ increases. This behavior reflects the fact that the effective tunnel coupling between the two dots weakens away from the symmetric configuration (see Supplementary Materials Section 4 for details). In contrast, the spin-flip regime (See Figs.~\ref{fig4}(b), (d), and (f)) exhibits qualitatively different behavior as $\lambda_R$ increases. Here, the Rabi dynamics are governed by the avoided crossing between opposite-spin states, enabled by the RSOC. A characteristic pointed region with enhanced oscillation amplitude emerges near $\Delta \delta$ $\sim$ $\pm\delta_0$, where the interdot spin-flip hybridization is resonantly enhanced.

The dependence on the RSOC strength further distinguishes the two regimes. In the spin-conserving case [Figs.~\ref{fig4}(a), (c), and (e)], the Rabi period remains nearly unchanged as the RSOC strength increases. Although $\Delta_{sc}$ decreases with increasing $\lambda_R$ as shown in Fig.~3(b), its effect on the Rabi dynamics is weak. This is because the effective Hamiltonian considered here does not account for the RSOC-induced deformation of the DQD potential.
Conversely, in the spin-flip regime (for Figs.~\ref{fig4}(b), (d), and (f)) , the Rabi period decreases with increasing RSOC. Specifically, Fig.~\ref{fig4}(b) shows the absence of Rabi oscillations in the spin-flip regime, which arises from the lack of an RSOC-induced energy gap.

This trend is consistent with the linewidth evolution shown in Supplementary Figure S4. As RSOC increases, the spin-conserving transition exhibits broader linewidths, whereas the spin-flip transition becomes sharper. 
\newline

\section{Discussion}
\label{discussion}
We now discuss the implications of RSOC in our system. The enhanced RSOC in our system places the spin splitting in an energy range that is both experimentally accessible and tunable in graphene-based spin qubits~\cite{Dulisch2025}. 
Although the resulting meV-scale spin splitting exceeds the typical microwave frequencies used in conventional semiconductor or NV-center qubits, it opens the possibility of ultrafast quantum operations. 
In practice, the effective driving frequencies can be engineered by tuning the RSOC strength, detuning, and magnetic field to match experimentally feasible control protocols.

Importantly, the RSOC strength provides a direct tuning knob for controlling the energy splitting in strained graphene qubits. Since the energy splitting is closely related to the spin relaxation time, RSOC tuning enables control over coherence properties and device performance. This highlights the potential of strained graphene as a platform where both mechanical strain and SOC can be co-engineered for quantum information applications.

We next discuss the readout mechanism of the proposed spin qubit. The qubit state can be detected via capacitive coupling to a nearby charge-sensitive element. Two feasible architectures can be considered: (i) a lateral multi-NB geometry, where a neighboring NB acts as a charge sensor, and (ii) a vertical heterostructure design in which the strain-engineered SLG is coupled to an additional sensing layer (e.g., graphene) through an insulating spacer such as hBN. These approaches are analogous to established readout schemes in superconducting and semiconductor qubits and provide a realistic, non-invasive route for state detection~\cite{Pashkin2003,Plastina2003,Yun2023,Petersson2010,Hecker2025,Banszerus2021}.

Finally, we comment on scalability toward multi-qubit operations. While the present work focuses on single-qubit control, the platform naturally extends to multi-qubit architectures via QH edge channels, which act as long-range coherent coupling buses between spatially separated NB qubits. This removes the need for direct wavefunction overlap, relaxes fabrication constraints, and enables modular qubit layouts, analogous to cavity-mediated coupling in superconducting systems. Since inter-qubit interactions are mediated by long-range QH coupling rather than short-range exchange, this approach offers a viable and scalable pathway for graphene-based spin qubits. A detailed study of two-qubit gate operations remains an important direction for future work.

\section{Conclusion}
\label{conclusion}
We have developed a spin-qubit platform in strain-engineered SLG by combining pseudomagnetic confinement with electrically tunable RSOC and Zeeman fields. 
Quantum transport simulations and an analytical four-band model revealed two characteristic avoided crossings--spin-conserving and spin-flip gaps--that enable mode-selective qubit manipulation governed by the strength of RSOC. The detuning-dependent Rabi dynamics further demonstrate controllable transitions between these operational regimes, establishing a unified mechanism in which strain-induced confinement, SOC, and Zeeman coupling collectively enable spin control within a single device architecture.

While BLG provides a convenient route for electrostatic confinement, strain-engineered SLG offers complementary advantages, including the preservation of intrinsic electronic properties and enhanced tunability via strain, Zeeman fields, and proximity-induced RSOC.
The realization of spin-qubit operation in SLG represents a notable conceptual advancement, as its gapless electronic structure has traditionally posed challenges for qubit implementation and has therefore directed most prior efforts toward bilayer systems.
By leveraging strain-induced confinement to form DQDs and by enabling controlled spin-state hybridization through externally induced SOC, our results establish SLG as a viable platform for spin qubits.
Furthermore, the demonstrated tunability via electrical detuning, magnetic fields, and externally controlled RSOC provides a flexible and experimentally accessible parameter space, which is essential for the robust operation and control of spin qubits in graphene-based devices. 

Overall, these findings indicate that strain-engineered graphene can support a new class of high-coherence and electrically addressable spin qubits, opening opportunities for scalable quantum architectures based on gapless two-dimensional materials.


\section*{Declarations}
The authors declare that they have no competing interests and there are no conflicts of interest.

\section*{Acknowledgement}
This work was supported by the Global-Learning \& Academic research institution for Master's$\cdot$PhD students, and Postdocs (LAMP) Program of the National Research Foundation of Korea (NRF) funded by the Ministry of Education (No. RS-2023-00285353) and the NRF (Nos. RS-2025-00557045, RS-2025-25436094), and the KISTI supercomputing center (No. KSC-2024-CRE-0521). The authors are grateful to Prof. Hee Chul Park for his insightful discussions and contributions to this work.

\bibliographystyle{cip-v3-bst-submit}
\bibliography{references_251030}

\section*{Appendix}
\subsection*{Quantum Transport Simulation}
We perform quantum transport simulations using the KWANT package, a Python library designed for tight-binding calculations with a focus on mesoscopic transport~\cite{kwant}. The system Hamiltonian is expressed in a tight-binding form as  

\begin{align}
H = & \sum_{i} \psi_i^\dagger \left( u_i \sigma_0 + \Delta_z \sigma_z \right)\psi_i \nonumber \\
&+ \sum_{\langle i,j \rangle} \psi_i^\dagger \Big[ t_{ij}^{\text{eff}} \sigma_0 
+ i\lambda_R \left( \vec{\sigma} \times \hat{d}_{ij}\right)_z \Big]\psi_j + \text{h.c.},
\end{align}
where $\psi_i^{\dagger} = (c_{i\uparrow}^{\dagger}, c_{i\downarrow}^{\dagger})$ is the creation operator at site $i$, $\sigma_{x,y,z}$ are Pauli matrices, $\Delta_z$ represents the Zeeman energy due to an external magnetic field, and $\lambda_R$ is the RSOC strength. The second term describes the modified nearest-neighbor hopping processes. The strain field alters the integrals of the hopping via a bond stretching mechanism, which we incorporate through the effective hopping parameter $t_{ij}^{\text{eff}}=t_0exp[-\beta(\frac{d_{ij}}{a_0}-1)]$, where $\beta\sim$3.37~\cite{Pereira2009}, $d_{ij}$ is the distance between the sites $i$ and $j$, $a_0$ is the lattice constant of 2.46 \AA, and the unstrained hopping amplitude is set as $t_0$ = 2.7 eV.  
\newline





\end{document}


\maketitle
\newpage

\section{Snake-like trajectory}
\label{trajectory}

\begin{figure*}[h]
 	\centering
 	\includegraphics[width=0.9\columnwidth]{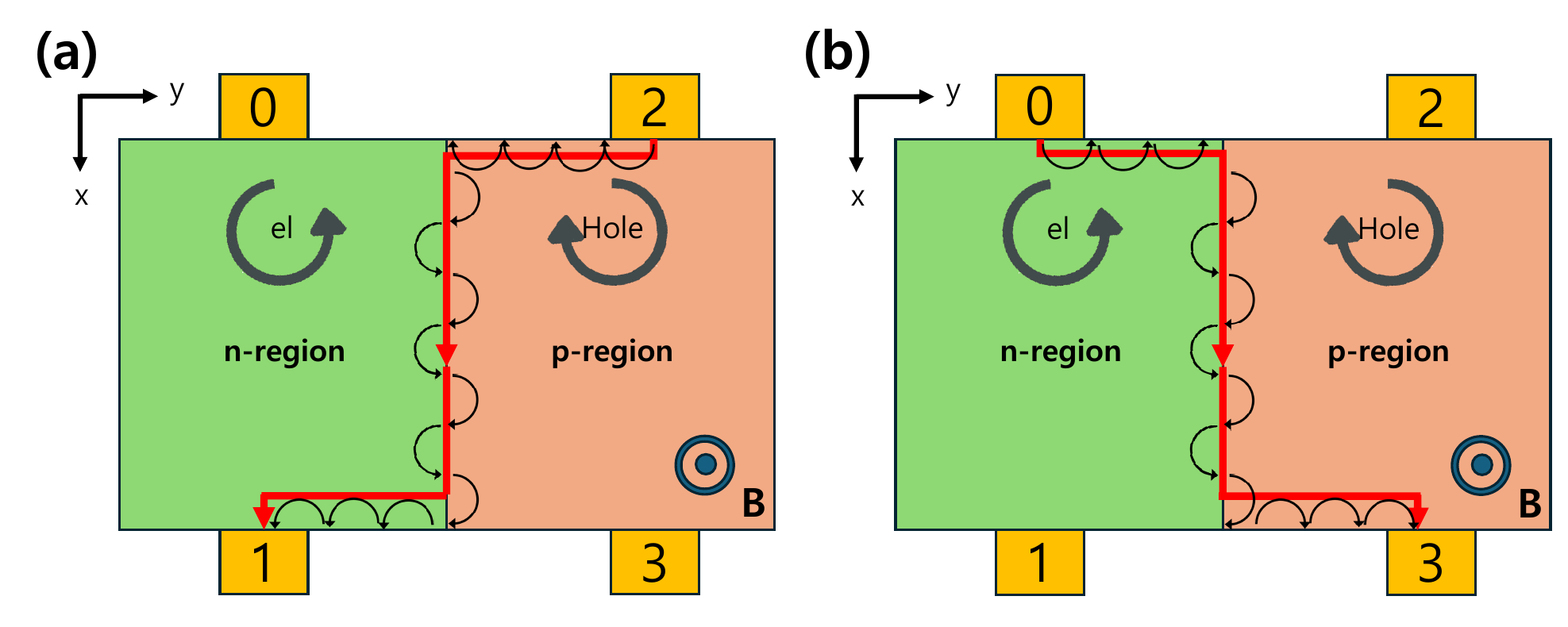}
 	\caption{\label{FigS1} Schematic illustration of current trajectory in a $p$-$n$ junction under a homogeneous perpendicular magnetic field. The panels (a) and (b) show possible configurations of quantum Hall edge channels along the junction interface. Electrons in the $n$-region and holes in the $p$-region propagate along chiral trajectories, forming snake-like states at the junction. The red arrows indicate representative current paths.
    }
\end{figure*}

The formation of such trajectories at a graphene $p$-$n$ junction under a perpendicular magnetic field originates from the opposite cyclotron motion of electrons and holes. In the $n$-region, electrons undergo cyclotron motion in one direction (e.g., counterclockwise), whereas in the $p$-region, holes bend in the opposite direction (clockwise), as illustrated in Supplementary Fig.~\ref{FigS1}. As a result, carriers crossing the interface experience alternating curvature, giving rise to snake-like trajectories that propagate along the junction.

In our system, we consider an antisymmetric $p$-$n$ junction with opposite Chern numbers in the two regions~\cite{Myoung2017}. In this case, the semiclassical picture naturally supports the formation of snake-like interface modes. These modes correspond to the transport path from lead 2 to lead 1, as shown in Supplementary Fig.~\ref{FigS1}(a), which is consistent with Fig. 1(a) in the main text. The corresponding transport path from the $n$-region to the $p$-region is illustrated in Supplementary Fig.~\ref{FigS1}(b). Importantly, when transport is considered in the opposite direction (from lead 0 to lead 3), the system exhibits qualitatively identical behavior, confirming the robustness of the snake-state transport.

\newpage
\section{Quantum conductance and Fano resonance line}
\label{quantum_cond}

\begin{figure*}[h]
 	\centering
 	\includegraphics[width=0.9\columnwidth]{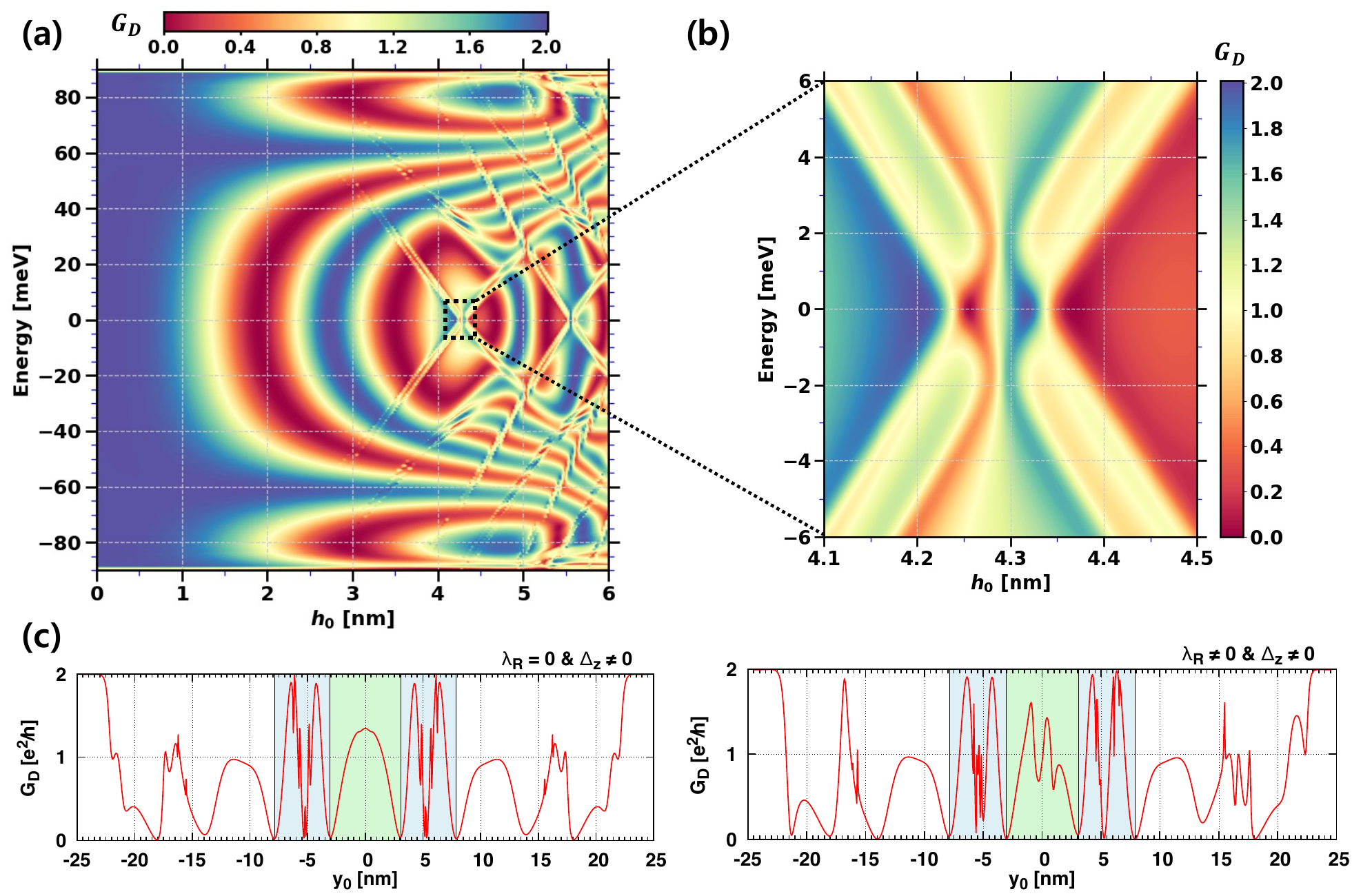}
 	\caption{\label{FigS2} Quantum conductance map and quantum conductance linecut at $h_0=4.3~nm$. 
    (a) Quantum conductance under only the Zeeman field ($1.7~meV$) without RSOC. 
    (b) The zoomed-in views around the single quantum dot (SQD) ground level in the range of $h_{0}=4.1$ to $4.5$ nm. 
    (c,d) Quantum conductance line as a function of $y_0$ for Zeeman coupling only (c) and for both RSOC ($10.1~meV$) and Zeeman coupling (d).
    }
\end{figure*}

Figure \ref{FigS2}(a) shows the conductance resonances without Rashba spin-orbit coupling (RSOC) in the quantum conductance $G_{D}$ characteristics of the strained graphene system, which are analogous to a spinful counterpart of those observed in the charge qubit system~\cite{Myoung2020,Park2023}. In the absence of RSOC, the Zeeman field produces two distinct spin-split branches, leading to independent crossing Fano resonance lines.
In contrast, when RSOC is taken into account, the spin-orbit interaction mixes the spin states.
The inclusion of RSOC modifies the conductance resonances in the strained graphene qubit system, as shown in Fig. 2(a) of the main text. 
The zoomed-in regions around the single quantum dot (SQD) regime are highlighted in Supplementary Fig. \ref{FigS2}(b) for the case without RSOC, which clearly exhibit the gap evolution near the ground-state level.

Supplementary Figures \ref{FigS2}(c) and \ref{FigS2}(d) display the line cuts of the quantum conductance $G_{D}$ as a function of the interface channel position $y_{0}$ for $h_0$=4.3 nm. Supplementary Figure \ref{FigS2}(c) considers Zeeman coupling only, while Supplementary Fig. \ref{FigS2}(d) includes both RSOC and Zeeman coupling.
Also, $G_D$ with RSOC clearly shows an asymmetric shape. 
The green area is the SQD activation region and the blue area is the DQD activation region, as shown in the main text, respectively.
\newpage

\section{Scalable tight-binding model}
\begin{figure*}[h]
 	\centering
 	\includegraphics[width=\columnwidth]{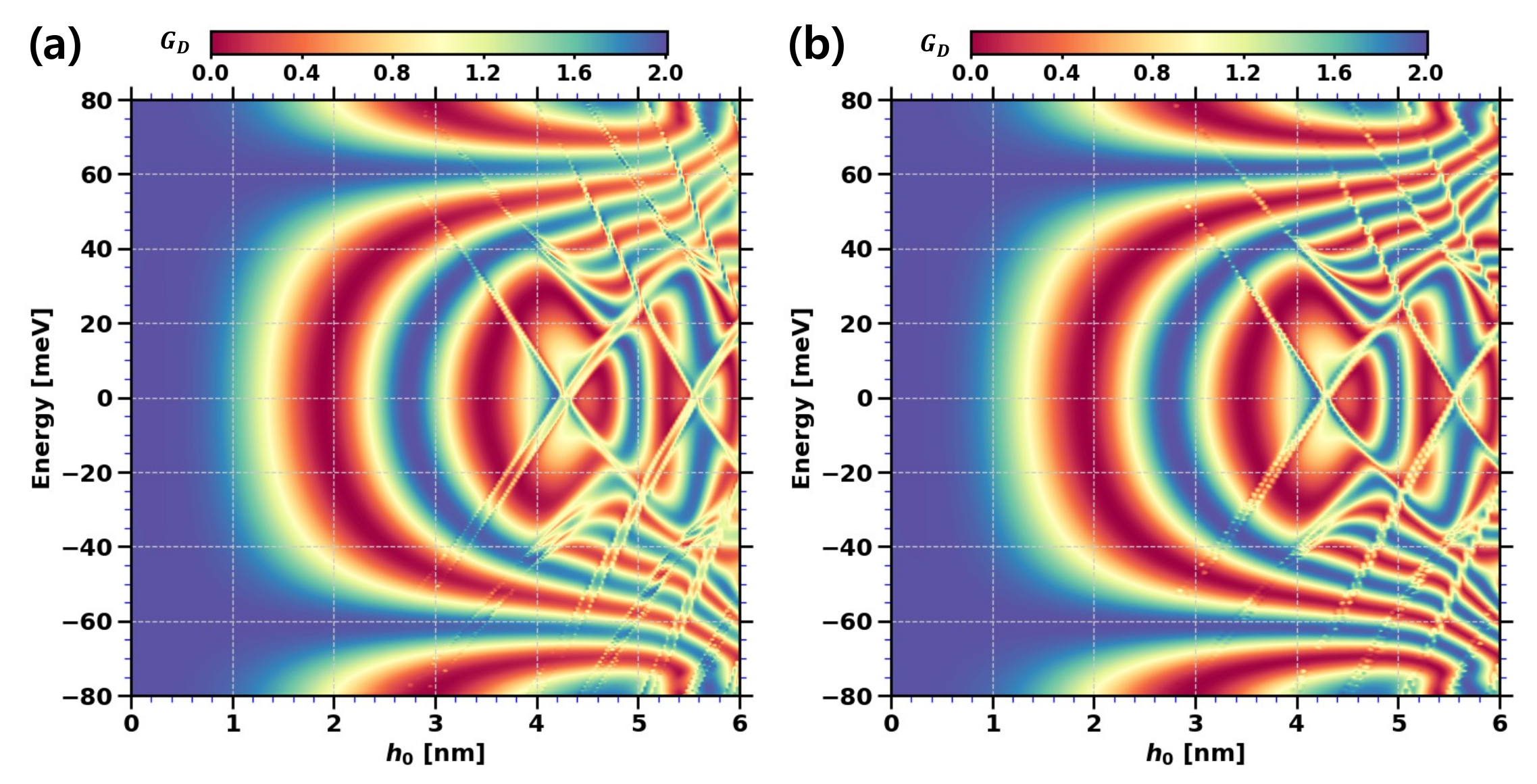}
 	\caption{\label{FigS3} 
    Quantum conductance $G_D$ maps as a function of energy and nanobubble height $h_0$. 
    The map for the original system (a) and a scaled system (b) for $\lambda_R=10.1~meV$ and $\Delta_z=1.7~meV$. 
    }
\end{figure*}
In the present calculations, a relatively large external magnetic field ($B_{z}\sim30~T$) is employed. This choice originates from the limited system size accessible in our numerical simulations, where the magnetic length and the confinement length scale are intrinsically linked. As a result, smaller simulated devices require larger magnetic fields to reproduce the same quantum Hall–assisted confinement physics.

Importantly, this does not imply that the proposed qubit mechanism intrinsically requires such high magnetic fields. For larger device sizes, the same physical regime can be achieved at proportionally smaller $B_{z}$.
This scaling behavior is consistent with recent studies~\cite{Liu2015,Liu2026} on strained graphene systems, where an efficient scaling approach has been demonstrated. Following this idea, we apply the transformation $a \rightarrow sa$, $t \rightarrow t/s$, $u \rightarrow su$, $h \rightarrow \sqrt{s}h$, and confirm that for scaling factor $s=2$, essentially identical results are obtained with a reduced magnetic field of $B_{z} \sim 7.5~T$.

These results indicate that the large magnetic field used in this work should be understood as a computational constraint rather than a fundamental requirement. Therefore, the spin-qubit behavior predicted here is expected to be realizable under more experimentally accessible magnetic fields in appropriately scaled strained-graphene devices.

\newpage
\section{Detuning energy versus RSOC}
\label{detuing_spectra}

\begin{figure*}[h]
 	\centering
 	\includegraphics[width=\columnwidth]{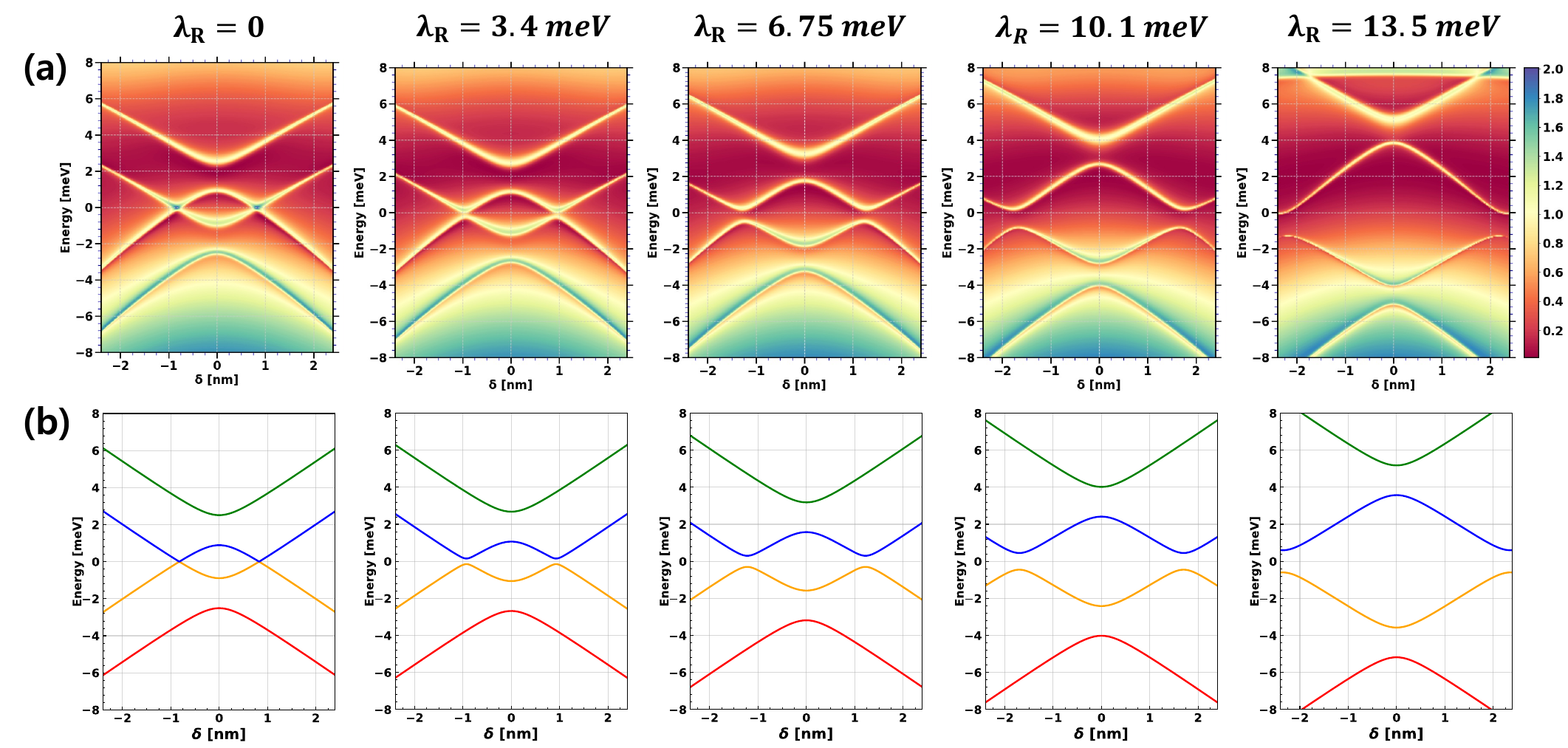}
 	\caption{\label{FigS4} 
    Detuning energy spectra as a function of $\lambda_R$. 
    (a) Numerical detuning spectra obtained from tight-binding simulations. 
    (b) Analytical detuning spectra calculated using the four-band Hamiltonian model.
    }
\end{figure*}

To clarify the role of RSOC in the energy spectrum as a function of detuning, we systematically investigated the energy spectral evolution by varying the RSOC strength under a fixed Zeeman splitting $\Delta_z=1.7~meV$. Supplementary Figure~\ref{FigS4}(a) presents the calculated energy spectra for increasing RSOC strengths ranging from $\lambda_R = 0$ to $13.5~meV$.

In the absence of RSOC ($\lambda_R = 0$), a finite spin-conserving gap is present, while the spin-flip gap remains closed. Upon introducing a finite RSOC, the spin-conserving gap is gradually reduced, reflecting a reshaping of the effective DQD confinement potential and a redistribution of electronic probability density between the two dots induced by RSOC (see ~\ref{eff_pot} for details). In contrast, the spin-flip gap increases approximately linearly with RSOC strength, as shown in Supplementary Fig. \ref{FigS6}(a). 

The linewidth of each energy level is related to the spin relaxation time $T_1$, following an inverse relation $\Gamma \propto 1/T_1$. As the RSOC strength increases, the linewidth associated with the upper spin-conserving branch becomes broader, indicating a reduction of the relaxation time due to enhanced spin-orbit-induced relaxation. Conversely, the linewidth of the spin-flip branch becomes progressively sharper, corresponding to a longer relaxation time.

We additionally analyzed the detuning spectrum using an effective four-band Hamiltonian model. The analytical results shown in Fig. \ref{FigS4}(b) are fully consistent with our numerical simulations.

Overall, RSOC provides a highly tunable mechanism to activate and control the detuning-induced spin hybridization. The progressive opening of the avoided crossing gap with increasing $\lambda_R$ establishes a direct link between RSOC strength, effective potential deformation, modification of the wavefunction overlap, and the emergence of electrically driven spin-flip Rabi dynamics, as experimentally probed and discussed in the main text.

\newpage
\section{Eigenstates and current density of detuning spectrum}
\label{eigenstate}

\begin{figure*}[h]
 	\centering
 	\includegraphics[width=\columnwidth]{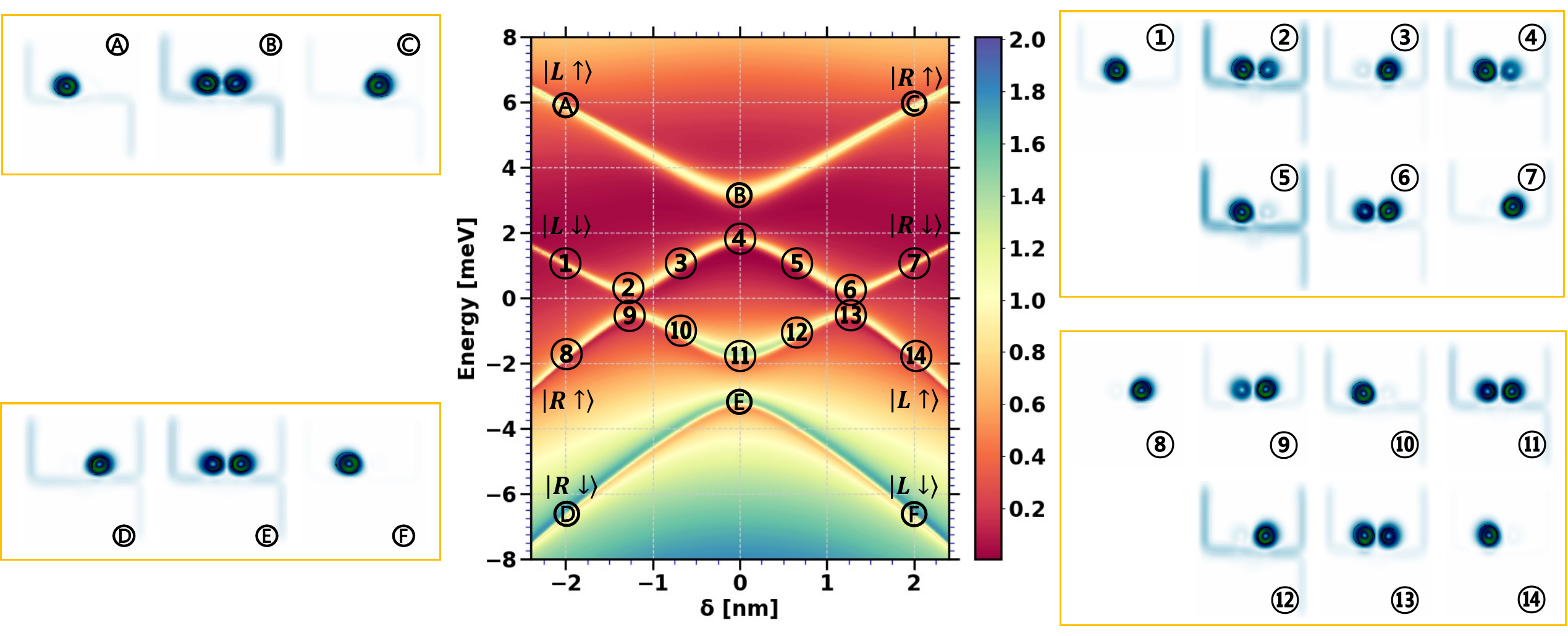}
 	\caption{\label{FigS5}
    Detuning-dependent energy spectrum and spatial localization of the eigenstates ($\lambda_R=6.75~meV$ and $\Delta_z=1.7~meV$).
    Central panel: Energy spectrum of the double quantum dot as a function of detuning $\delta$, showing the four hybridized states labeled as $|L\uparrow\rangle$, $|L\downarrow\rangle$, $|R\uparrow\rangle$, and $|R\downarrow\rangle$. 
    The colored circles mark the eigenenergies at selected detuning values.
    Left and right panels: Current density corresponding to the numbers and alphabet labels in the central spectrum.
    }
\end{figure*}

Supplementary Figure \ref{FigS5} illustrates the detuning energy spectrum of the graphene DQDs, displaying the four relevant eigenvalues.
Each colored circle represents the current density of the corresponding eigenstate evaluated at a specific energy and detuning value, providing a direct visualization of the real-space localization and hybridization of the quantum states (without explicitly showing spin states).

The left and right quantum dots are denoted as $\ket{L}$ and $\ket{R}$, respectively. From top to bottom, the four eigenstates are labeled as
$\ket{L\uparrow}$, $\ket{L\downarrow}$, $\ket{R\uparrow}$, and $\ket{R\downarrow}$, 
corresponding to the spin-resolved localized states of each dot.

As shown in Supplementary Fig.~\ref{FigS5}, near the avoided crossing point the eigenstates hybridize into superposed states across the two dots, leading to spatial delocalization of the wavefunctions.
At zero detuning ($\delta = 0$), the superposition occurs between states with the same spin orientation, resulting in bonding and antibonding superpositions of the form
\begin{equation}
\ket{\psi}_{upper} = \frac{1}{\sqrt{2}}
\left( \ket{L,\uparrow} \pm \ket{R,\uparrow} \right),
\quad
\text{or}
\quad
\ket{\psi}_{lower} = \frac{1}{\sqrt{2}}
\left( \ket{L,\downarrow} \pm \ket{R,\downarrow} \right),
\end{equation}
which correspond to the spin-conserving qubit operation regime.

In contrast, at the detuning positions $\delta = -\delta_0$ and $\delta = +\delta_0$, avoided crossings occur between opposite-spin states due to the RSOC. 
These crossings give rise to spin-hybridized superposition states
\begin{align}
\delta = -\delta_0 &: \quad 
\ket{\psi} = \frac{1}{\sqrt{2}}
\left( \ket{L,\downarrow} \pm \ket{R,\uparrow} \right),\\[4pt]
\delta = +\delta_0 &: \quad 
\ket{\psi} = \frac{1}{\sqrt{2}}
\left( \ket{L,\uparrow} \pm \ket{R,\downarrow} \right),
\end{align}
which serve as the basis states for electrically driven spin-flip qubit manipulations.

These eigenstate visualizations confirm that the qubit operation regimes identified in the detuning spectrum originate from state-selective hybridization between the DQDs, controlled by the combined effect of detuning and RSOC.

\newpage
\section{Deformation of tunnel coupling}
\label{eff_pot}

\begin{figure*}[h]
 	\centering
 	\includegraphics[width=0.7\columnwidth]{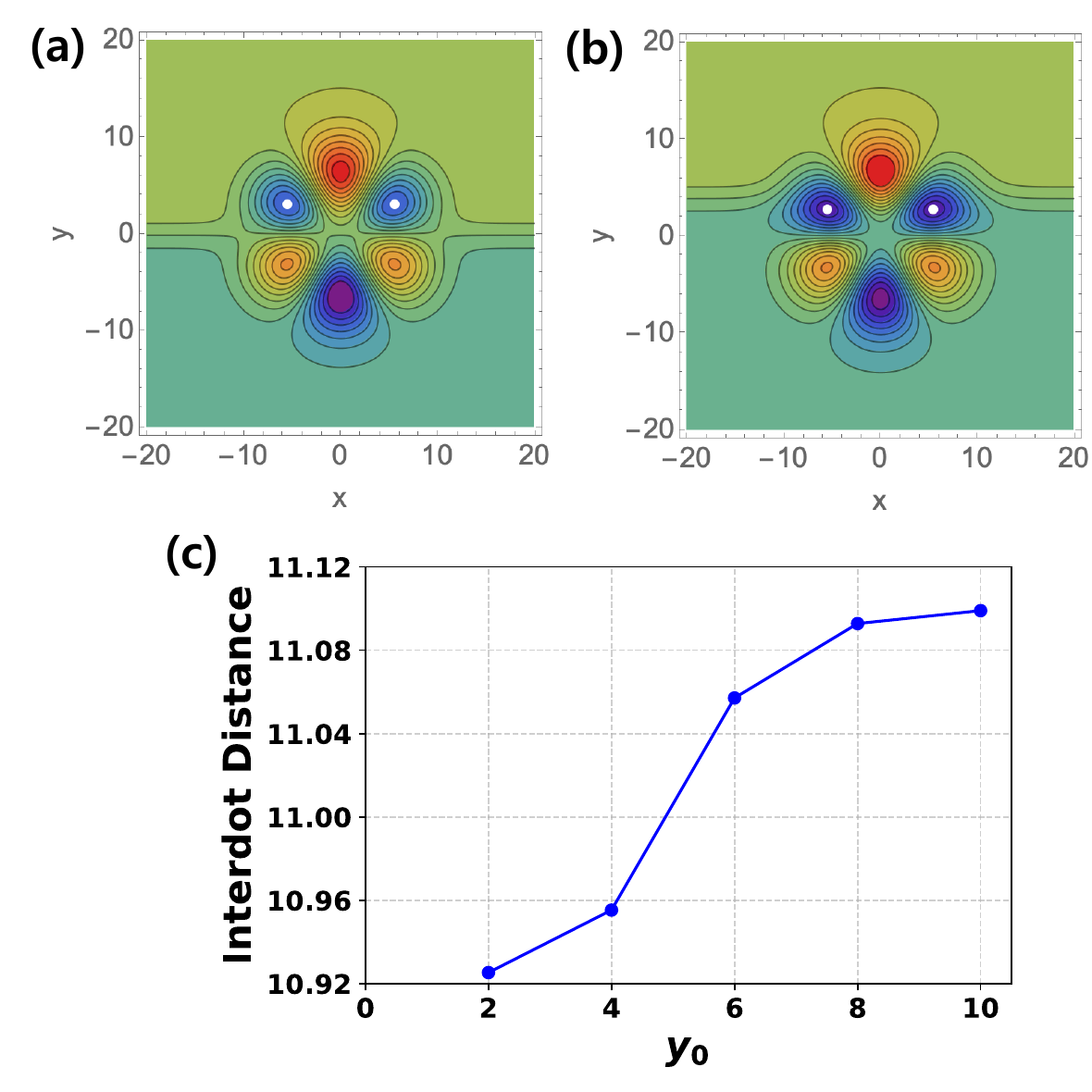}
 	\caption{\label{FigS6} 
    Effective potential landscape and interdot distance as a function of $y_0$.
    (a,b) Two-dimensional maps of the effective potential $V_{\mathrm{eff}}$ for $y_0 = 0$ (a) and $y_0 = 6$ (b), where $y_0$ denotes the position of the $p$-$n$ interface channel. The contour plots illustrate the evolution of the two local potential minima defining the DQDs. 
    (c) Interdot distance between the minimum points of two QDs as a function of $y_0$.
    }
\end{figure*}

As mentioned in the main text, the $\Delta_{\mathrm{sc}}$ is not explicitly related to RSOC. Nevertheless, the gap decreases with increasing RSOC strength, as shown in Fig. 3(b) in the main text. To gain further insight into these gap behaviors, we analyze the gap characteristics using the effective potential.
The effective potential is written as Equation (2) in the main text.

Supplementary Figures~\ref{FigS6}(a) and (b) show how the effective potential landscape evolves as the junction interface position $y_0$ is shifted. The blue region denotes the confining potential well, and the white dots mark the local minima associated with the two quantum dots. As the interface moves toward positive $y$, the potential well becomes increasingly deformed, which in turn enlarges the spatial separation between the two local minima. This geometric modification reduces the overlap between the localized wave functions, effectively weakening the interdot coupling.

The consequence of this deformation is quantified in Supplementary Fig.~\ref{FigS6}(c), where the increasing minimum-to-minimum distance correlates with a reduction in the measured spin-conserving gap $\Delta_{\mathrm{sc}}$. Importantly, the intrinsic $\Delta_{\mathrm{sc}}$ itself is not directly altered by the RSOC strength; rather, the apparent change arises because increasing the RSOC strength induces an effective displacement of the $p$-$n$ junction interface, which subsequently modifies the DQD confinement potential. In this sense, RSOC tuning mimics the effect of shifting $y_0$, providing a unified interpretation of how RSOC indirectly influences $\Delta_{\mathrm{sc}}$ through deformation of the effective potential landscape.

\newpage
\section{Kramers degeneracy}
\label{Kramer}
To examine the symmetry of the effective Hamiltonian in Eq.~(3) in the main text, we express it in terms of Pauli matrices in the basis 
$\{ |L,\uparrow\rangle,\ |L,\downarrow\rangle,\ |R,\uparrow\rangle,\ |R,\downarrow\rangle \}$.
The Hamiltonian projected onto this subspace is given by
\begin{equation}
\label{equationS4}
    H_K(\delta,\Delta_z,\lambda_R,t_c) 
    = \frac{\alpha\delta}{2}\tau_z\otimes\sigma_0 
    + t_c\tau_x\otimes\sigma_0 
    + \frac{\Delta_z + \gamma\lambda_R^2}{2}\tau_0\otimes\sigma_z 
    + \frac{\beta\lambda_R}{2}\tau_y\otimes\sigma_y ,
\end{equation}
where $\tau_i$ and $\sigma_i$ denote the Pauli matrices acting on the dot ($L/R$) and spin subspaces, respectively.

We now examine the time-reversal symmetry (TRS) of this Hamiltonian. 
The time-reversal operator for spin-$1/2$ systems is given by
\begin{equation}
\mathcal{T} = I_{\tau} \otimes (i\sigma_y) K,
\end{equation}
where $K$ denotes complex conjugation. 
Under time reversal, the spin Pauli matrices transform as $\boldsymbol{\sigma} \rightarrow -\boldsymbol{\sigma}$, while the orbital (dot) Pauli matrices remain unchanged, $\boldsymbol{\tau} \rightarrow \boldsymbol{\tau}$, and $i \rightarrow -i$ due to complex conjugation.

The detuning term $\tau_z \otimes \sigma_0$ and the tunneling term $\tau_x \otimes \sigma_0$ are spin-independent and therefore invariant under $\mathcal{T}$. 
The RSOC term $\tau_y \otimes \sigma_y$ also preserves TRS, since $\sigma_y$ changes sign under time reversal while $\tau_y$ remains unchanged, leaving the product invariant. 
In contrast, the Zeeman term $\tau_0 \otimes \sigma_z$ changes sign under time reversal and therefore breaks TRS. 
We note that this Hamiltonian describes only a single-valley subspace of the full system, and thus does not by itself need to preserve TRS.

We now consider the Hamiltonian of the total system including both valleys. 
In the basis $\{ |K,L,\uparrow\rangle,\ |K,L,\downarrow\rangle,\ |K,R,\uparrow\rangle,\ |K,R,\downarrow\rangle, 
|K',L,\uparrow\rangle,\ |K',L,\downarrow\rangle,\ |K',R,\uparrow\rangle,\ |K',R,\downarrow\rangle \}$, 
and in the absence of intervalley scattering, the Hamiltonian can be written as
\begin{equation}
\label{equationS7}
    H_{\mathrm{tot}} = 
    \begin{bmatrix}
        H_K & 0 \\
        0 & H_{K'}
    \end{bmatrix},
\end{equation}
with
\begin{equation}
\label{equationS8}
    H_{K'}(\delta,\Delta_z,\lambda_R,t_c) 
    = \frac{\alpha\delta}{2}\tau_z\otimes\sigma_0 
    + t_c\tau_x\otimes\sigma_0 
    + \frac{\Delta_z - \gamma\lambda_R^2}{2}\tau_0\otimes\sigma_z 
    + \frac{\beta\lambda_R}{2}\tau_y\otimes\sigma_y.
\end{equation}

To describe the full system in a compact form, we introduce the Pauli matrices $\nu_i$ acting in the valley subspace. 
In the basis $(K,K')\otimes(L,R)\otimes(\uparrow,\downarrow)$, the total Hamiltonian is written as
\begin{equation}
\label{equationS9}
    H_{\mathrm{tot}} 
    = \frac{\alpha\delta}{2}\nu_0\otimes\tau_z\otimes\sigma_0 
    + t_c\nu_0\otimes\tau_x\otimes\sigma_0 
    + \frac{\Delta_z}{2}\nu_0\otimes\tau_0\otimes\sigma_z 
    + \frac{\gamma\lambda_R^2}{2}\nu_z\otimes\tau_0\otimes\sigma_z 
    + \frac{\beta\lambda_R}{2}\nu_0\otimes\tau_y\otimes\sigma_y ,
\end{equation}
where $\nu_i$ are Pauli matrices acting in the valley subspace. 
The second-order Rashba-induced term is valley-odd, whereas the other terms are valley-even.

Under time reversal, the spin Pauli matrices transform as $\boldsymbol{\sigma} \rightarrow -\boldsymbol{\sigma}$, while the valley index is exchanged ($K \leftrightarrow K'$), corresponding to $\nu_z \rightarrow -\nu_z$. 
As a result, the valley-odd term $\nu_z \otimes \sigma_z$ remains invariant under TRS, whereas the Zeeman term $\sigma_z$ changes sign. 
Therefore, the total Hamiltonian preserves TRS only when $\Delta_z = 0$, while any finite Zeeman field explicitly breaks TRS.

\newpage
\section{The effective 4-bands Hamiltonian of our device model}
\label{4bands_hamil}
\subsection{Second-order term of RSOC}
We note that the Rashba term does not directly split the spin states at first order but generates a second-order virtual energy shift, $\delta E \sim \gamma\lambda_R^2$, which appears as a Zeeman-like correction after projection onto the low-energy sector.

We decompose the Hamiltonian of Eq.~\eqref{equationS4} into an unperturbed part and a perturbation as
\begin{equation}
    H = H_0 + V
\end{equation}
where
\begin{equation}
\begin{array}{cl}
    H_0 = \frac{\alpha\delta}{2}\tau_z + t_c\tau_x,
    \\ \\
    V = \frac{\Delta_z}{2}\sigma_z
\end{array}
\end{equation}

The eigenstates of $H_0$ are written by $|s,\eta\rangle$, with corresponding eigenenergies
\begin{equation}
    E_{s,\eta}^{(0)} = s\Omega + \eta\frac{\Delta_z}{2}
\end{equation}
where $s=\pm1$ labels the isospin (bonding/antibonding), $\eta=\pm1$ denotes the spin, and 
\begin{equation}
    \Omega = \sqrt{(\frac{\alpha\delta}{2})^2 + t_c^2}.
\end{equation}

The first-order correction vanishes because the perturbation $V \propto \tau_y\sigma_y$ flips both the orbital and spin indices. Therefore, it does not couple a state to itself but only connects it to orthogonal states, leading to a zero diagonal matrix element,
\begin{equation}
    E_{s,\eta}^{(1)}=\langle s,\eta |V| s,\eta \rangle,
\end{equation}
since both $\tau_y$ and $\sigma_y$ are purely off-diagonal in their respective eigenbases.

Using standard nondegenerate perturbation theory, the second-order energy correction is given by
\begin{equation}
\label{equationS14}
    E_{s,\eta}^{(2)}= \displaystyle\sum_{(s',\eta')\neq(s,\eta)} \frac{|\langle s',\eta' |V| s,\eta \rangle|^2}{E_{s,\eta}^{(0)} - E_{s',\eta'}^{(0)}}.
\end{equation}

Since the perturbation $V=(\beta\lambda_R/2)\tau_y\sigma_y$ flips both the isospin and spin indices, the only nonzero matrix element connects $|s,\eta\rangle$ to $|-s,-\eta\rangle$. Thus, the summation reduces to a single term, yielding
\begin{equation}
    E_{s,\eta}^{(2)}= \frac{(\frac{\beta\lambda_R}{2})^2}{E_{s,\eta}^{(0)} - E_{s',\eta'}^{(0)}}.
\end{equation}

Evaluating the denominator explicitly, we obtain
\begin{equation}
    {E_{s,\eta}^{(0)} - E_{s',\eta'}^{(0)}} =2s\Omega + \eta\Delta_z.
\end{equation}

Therefore, the second-order correction takes the form
\begin{equation}
\label{equationS17}
    E_{s,\eta}^{(2)}= \frac{(\frac{\beta\lambda_R}{2})^2}{2s\Omega + \eta\Delta_z}.
\end{equation}

This result shows that the RSOC does not contribute at first order. As shown in Eq.~\eqref{equationS14}, the second-order process returns the system to its initial state through two successive perturbations, resulting in a diagonal energy renormalization via virtual spin-flip and interdot transitions. This mechanism is consistent with previous studies, where RSOC induces mixing of spin states and leads to level anticrossing through higher-order processes~\cite{Bulaev2005,Stano2007}.

\subsection{TRS analysis of the RSOC-induced coupling}

\begin{figure*}[ht]
 	\centering
 	\includegraphics[width=1.0\columnwidth]{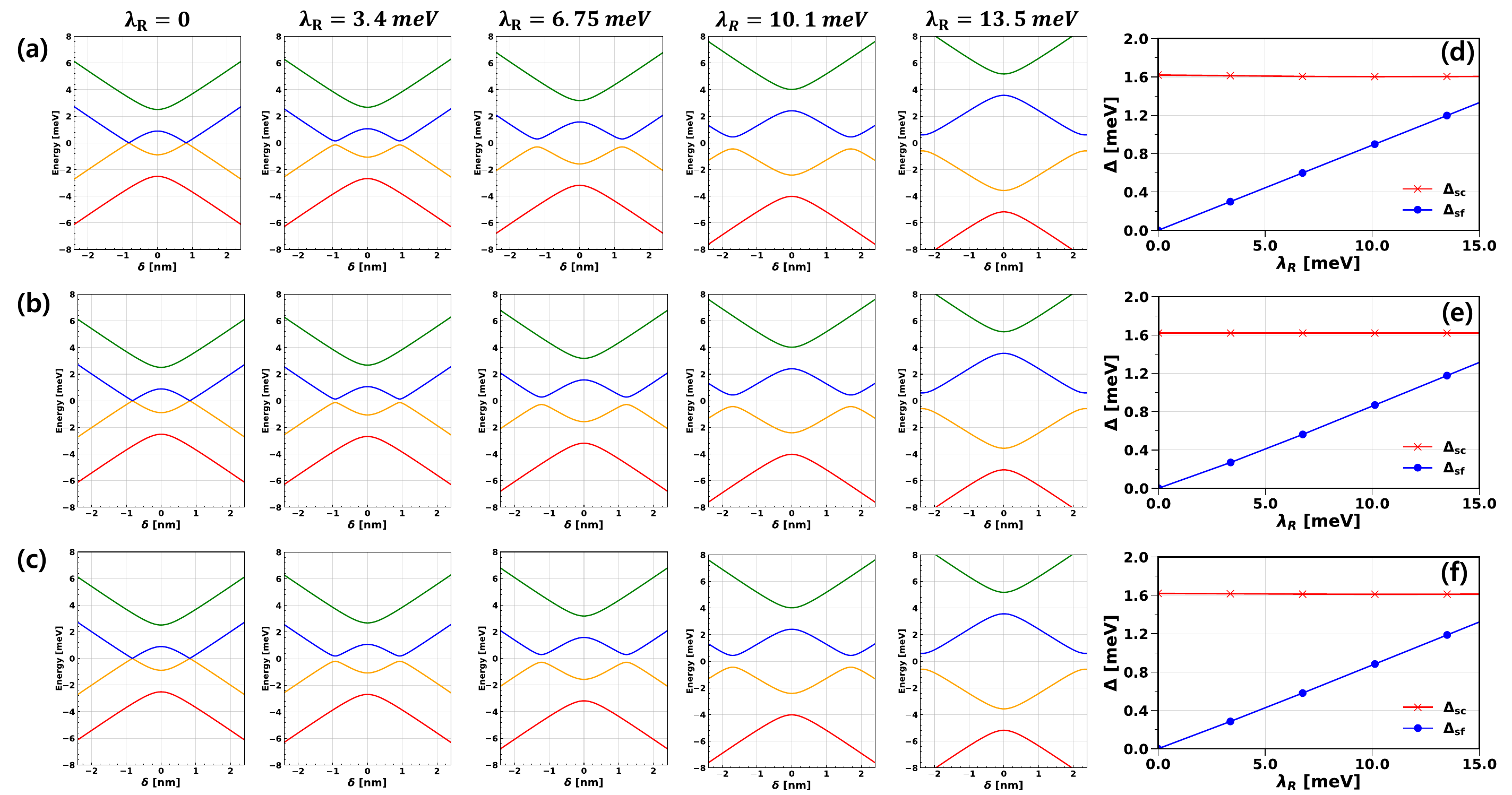}
 	\caption{\label{FigS7} 
    Comparison of effective-model fittings and gap evolutions for different RSOC coupling structures in the four-band Hamiltonian. Effective-model spectra generated using (a) the $\tau_y\sigma_y$ RSOC term, (b) the $\tau_x\sigma_y$ RSOC term, and (c) both coupling terms simultaneously.
    (d-f) Gap evolutions as a function of $\lambda_R$ at fixed $\Delta_z$, corresponding to (a), (b), and (c), respectively.}
\end{figure*}

To further examine the form of the RSOC-induced coupling in the projected four-band Hamiltonian, we compared effective models containing (i) the $\tau_y \sigma_y$ term, (ii) the $\tau_x \sigma_y$ term, and (iii) both coupling terms simultaneously. As shown in Fig.~\ref{FigS7}, all three models reproduce qualitatively similar detuning spectra and gap evolutions. Therefore, the choice of the minimal coupling term should additionally be guided by the symmetry properties of the projected low-energy model.

Since RSOC arises from relativistic spin-orbit interaction associated with inversion-symmetry breaking rather than magnetic ordering, the RSOC-induced coupling should remain compatible with the effective time-reversal-symmetric structure of the zero-field projected Hamiltonian. As discussed in ~\ref{Kramer}, under the time-reversal operation $T=\tau_0 \otimes i\sigma_y K$, the spin Pauli matrix changes sign, $\sigma_y \rightarrow -\sigma_y$, while $\tau_y$ also changes sign because it is purely imaginary. Consequently, the product $\tau_y \sigma_y$ remains invariant under time reversal and preserves the zero-field degeneracy structure of the projected model. In contrast, a $B$-independent $\tau_x \sigma_y$ term is odd under time reversal within the projected low-energy model, because $\tau_x$ remains unchanged whereas $\sigma_y$ changes sign under the time-reversal operation. Such a term would therefore explicitly break the effective time-reversal-symmetric structure of the zero-field Hamiltonian. We therefore adopt the $\tau_y \sigma_y$ term as the minimal symmetry-compatible representation of the RSOC-induced spin-flip coupling in our projected four-band Hamiltonian.

\subsection{Full matrix form of effective 4-bands Hamiltonian}
The full matrix representation of Eq.~\eqref{equationS4} is given by
\[
\scalebox{0.85}{$
H =
\begin{bmatrix}
\label{4band}
\dfrac{\alpha \delta}{2} + \dfrac{1}{2}\Bigl(\Delta_z + \gamma \lambda_R^{2}\Bigr) 
& 0 
& t_c\ 
& -\dfrac{1}{2}\,\beta \lambda_R \\[1.2em]
0 
& \dfrac{\alpha \delta}{2} - \dfrac{1}{2}\Bigl(+\Delta_z + \gamma \lambda_R^{2}\Bigr) 
& \dfrac{1}{2}\,\beta \lambda_R 
& t_c\ \\[1.2em]
t_c\ 
& \dfrac{1}{2}\,\beta \lambda_R 
& -\dfrac{\alpha \delta}{2} + \dfrac{1}{2}\Bigl(\Delta_z + \gamma \lambda_R^{2}\Bigr) 
& 0 \\[1.2em]
-\dfrac{1}{2}\,\beta \lambda_R 
& t_c\ 
& 0 
& -\dfrac{\alpha \delta}{2} - \dfrac{1}{2}\Bigl(+\Delta_z + \gamma \lambda_R^{2}\Bigr)
\end{bmatrix}
$}
\]

By fitting to the result of our tight-binding numerical simulation, we obtained the following parameters in the effective Hamiltonian: 
$t_c=0.3$, $\Delta_{z}=1.26$, $\lambda=2n\times\Delta_{z}$ ($n$=integer), $\alpha=0.33$, $\beta=0.022$, and $\gamma=0.0048$.

This effective four-band Hamiltonian enables us to analyze the evolution of the characteristic energy gaps systematically. In particular, we focus on the spin-conserving gap, the spin-flip gap, and their dependence on the key parameters. Specifically, we investigate how the gaps evolve as functions of Zeeman energy, RSOC strength, detuning, and interdot coupling.
Supplementary Figure ~\ref{FigS4}(b) presents the energy spectrum as a function of detuning, which corresponds to the same behavior shown in Supplementary Fig. ~\ref{FigS4}(a). From these spectra, we extracted the gap evolution using Eq.~\eqref{equationS4}, and the results are summarized in Supplementary Figure \ref{FigS7}(d).
The trend obtained by our effective four-band Hamiltonian is consistent with the numerical simulation results shown in Fig. 3(b) in the main text, confirming the validity of our four-band model.

\newpage
\section{Intervalley scattering}
\label{interval_scatt}

\begin{figure*}[h]
 	\centering
 	\includegraphics[width=0.9\columnwidth]{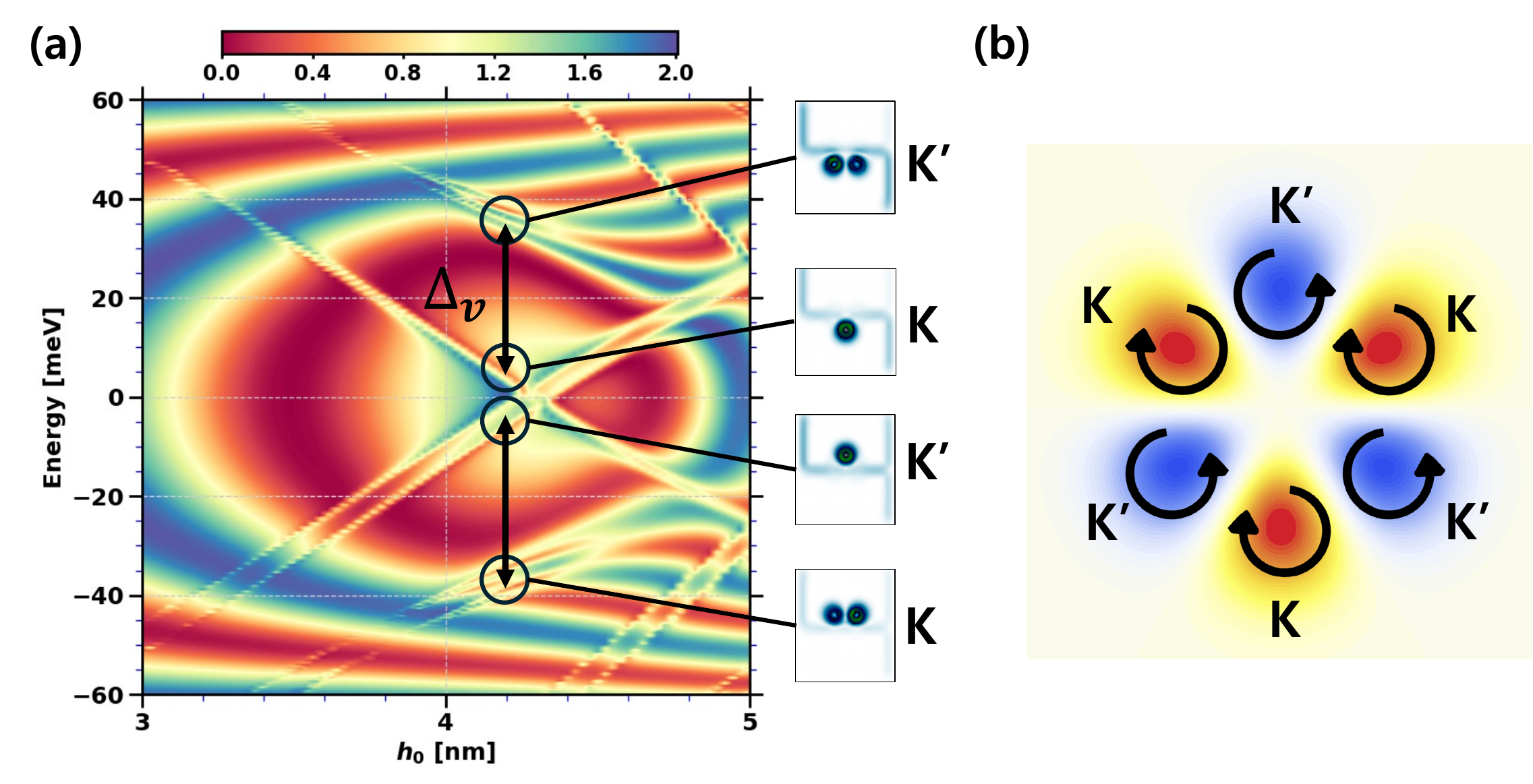}
 	\caption{\label{FigS8} 
    (a) Quantum conductance map as a function of nanobubble height $h_0$ and energy, with the corresponding current density distribution at selected Fano resonance lines shown on the right. $\Delta_{v}$ indicates the energy splitting between $K$ and $K'$ valley. 
    (b) Schematic illustration of valley-dependent cyclotron motion induced by PMF.}
\end{figure*}

Intervalley scattering in graphene requires a large momentum transfer on the order of $|K-K'|$, which can only be induced by atomically sharp disorder or short-range potentials. In contrast, the smoothly varying $p$-$n$ junction and Gaussian-shaped nanobubble considered in our system do not provide such momentum transfer, thereby suppressing intervalley mixing. 
Without spin degrees of freedom~\cite{Myoung2020}, the intervalley splitting $\Delta_v$ is approximately 40 meV, which is significantly larger than the relevant energy scales considered in our model, including the RSOC ($\lambda_R \sim 10.1~meV$) and the Zeeman energy ($\Delta_z \sim 1.7 ~meV$).
As a result, the qubit defined within a single valley is not affected by the opposite PMF in the other valley, and its coherence properties are expected to remain robust. Therefore, valley mixing is negligible in our system, justifying the single-valley description used throughout this work.

\newpage
\section{Calculation Methods}
\subsection*{Strain-induced Nanobubble}
The nanobubble is assumed to follow a Gaussian profile, which describes the out-of-plane displacement field,  
\begin{equation}
z(r) = h_0 \exp\!\left(-\frac{r^2}{2\sigma^2}\right),
\end{equation}
where $h_0$ denotes the maximum height of the bubble and $\sigma$ characterizes its lateral width. In this study, we use $h_0$ = 4.29 nm and $\sigma$ = 7.38 nm. 
Such nanobubbles are commonly observed in graphene placed on a substrate, arising from trapped molecules or strain relaxation~\cite{Jia2019,Ghorbanfekr2017,Zamborlini2015,Faraji2022,Villarreal2021,Nikolai2012,Zhu2014}. Local strain in nanobubbles naturally generates PMFs, which confine Dirac fermions and lead to the formation of localized quantum states, effectively acting as single- and double-quantum-dot states, as shown in Fig. 1(a) of the main text. 
To probe and manipulate these quantum-dot-like states, we place the nanobubble at the interface of a graphene $p$-$n$ junction. In the quantum Hall regime, the $p$-$n$ junction hosts chiral interface channels, which can couple to the localized states in the strained region. This coupling realizes a hybrid system, where quantum Hall edge currents interact with the PMF-generated quantum dots, forming a controllable two-level system (TLS) suitable for qubit operations.

\subsection*{Quantum Transport}
The transport properties are computed within the Landauer-B\"uttiker formalism. The conductance at energy $E$ is given by  
\begin{equation}
G(E) = \frac{2e^2}{h} T_{\alpha\beta}(E),
\end{equation}
where $T_{\alpha\beta}(E)$ is the transmission probability from lead $\beta$ to lead $\alpha$. This transmission is obtained from the scattering matrix $S$ of the system,  
\begin{equation}
T_{\alpha\beta}(E) = \sum_{n\in\alpha, m\in\beta} \left| S_{nm}(E) \right|^2,
\end{equation}
with propagating modes labeled $n_\alpha$ and $m_\beta$ in the corresponding leads. In our model, spin degeneracy is explicitly included so that the conductance spectra exhibit quantized steps in units of $e^2/h$ per total spin channel. This allows us to resolve the interplay between spin splitting, Rashba coupling, and pseudomagnetic confinement in the proposed nanobubble-based spin qubit system.

\section*{Rabi Oscillations}

\subsubsection*{Basis of the Four-band Hamiltonian}

Following the localized basis defined in Fig.~3(a) of the main text,
\[
\{|L,\uparrow\rangle,\;|L,\downarrow\rangle,\;|R,\uparrow\rangle,\;|R,\downarrow\rangle\},
\]
we use the four-band Hamiltonian given in Eq. (3) of the main text to perform Rabi oscillations.
As shown in the energy spectra in Supplementary Fig.~\ref{FigS4}, the detuning parameter
$\delta$ controls the electrostatic offset between the two dots and thereby
drives the transitions at the relevant avoided crossings.  
As discussed in Sec.~2.3 of the main text, these avoided crossings fall into
two categories: spin-conserving and spin-flip transitions.

\subsubsection*{Identification of the avoided crossings}

For each detuning value~$\delta$, we compute the four eigenvalues of
Eq. (3) of the main text, ordered as
\[
E_1(\delta) \le E_2(\delta) \le E_3(\delta) \le E_4(\delta).
\]
\paragraph{$\blacktriangleright$ spin-conserving avoided crossing.}
This avoided crossing corresponds to the upper pair of eigenvalues.
The gap is defined as
\begin{equation}
\Delta_{\mathrm{sc}}(\delta) = E_4(\delta) - E_3(\delta),
\end{equation}
and occurs at $\delta = 0$ for our model parameters.

\paragraph{$\blacktriangleright$ spin-flip avoided crossing.}
The Rashba spin-orbit-mediated avoided crossing involves the middle pair of
eigenvalues, with the gap
\begin{equation}
\Delta_{\mathrm{sf}}(\delta) = E_3(\delta) - E_2(\delta).
\end{equation}
The detuning point $\delta_0 < 0$ is obtained by minimizing
$\Delta_{\mathrm{sf}}(\delta)$ over negative~$\delta$.

These avoided crossings define the relevant two-level subspaces for the
Rabi-map simulations.

\subsubsection*{Observables for the two interdot transitions}

Detuning controls two distinct interdot transitions in the DQD, as explained in Section~2.3 in the main text.  
Because the two transitions involve different pairs of states, we employ different observables for the dynamics.

\paragraph{$\blacktriangleright$ spin-conserving interdot transition.}
When both participating states share the same spin orientation
(e.g., $|L,\uparrow\rangle$ and $|R,\uparrow\rangle$), detuning induces an interdot charge transfer without flipping the spin as shown in Fig. 1(c) of the main text.  
The appropriate observable is 
\begin{equation}
\tau_z
= (|L,\uparrow\rangle\langle L,\uparrow|
   + |L,\downarrow\rangle\langle L,\downarrow|)
 - (|R,\uparrow\rangle\langle R,\uparrow|
   + |R,\downarrow\rangle\langle R,\downarrow|).
\end{equation}
This quantity measures the left-right charge transfer and corresponds to the observable denoted as $\langle \tau_z\rangle$ in Figs.~4(a), (c), and (e) of the main text.

\paragraph{$\blacktriangleright$ spin-flip interdot transition.}
The spin-flip transition occurs between $|L,\downarrow\rangle$ and
$|R,\uparrow\rangle$, where spin-orbit coupling mediates a combined interdot tunneling between different spin orientations as shown in Fig. 1(d) of the main text.
To capture this process, we use the population difference between the two states forming the avoided crossing,
\begin{equation}
\sigma_z
= |L,\downarrow\rangle\langle L,\downarrow|
 - |R,\uparrow\rangle\langle R,\uparrow|.
\end{equation}
This observable captures the spin-orbit-assisted interdot transition and is labeled $\langle \sigma_z\rangle$ in Figs.~4(b), (d), and (f) of the main text.

\subsubsection*{Liouvillian time evolution}

The density matrix obeys the Lindblad master equation
\begin{equation}
\frac{d\rho}{dt}
= -i[H,\rho] + \sum_k (L^{\dagger}_k \rho L_k - \frac{1}{2} \{(L^{\dagger}_k L_k, \rho\}.
\end{equation}

Relaxation and dephasing within the two-level subspace are included through the collapse operators
\begin{equation}
\Gamma_1 = \sqrt{1/T_1}\,|g\rangle\langle e|,
\qquad
\Gamma_\phi = \sqrt{1/T_2 - 1/(2T_1)}
      \left(|e\rangle\langle e| - |g\rangle\langle g|\right),
\end{equation}
where $|e\rangle$ and $|g\rangle$ denote the relevant pair of states for each type of interdot transition.

We solve the master equation $\dot{\rho}=L\rho$ by computing the exponential action $e^{Lt}|\rho(0)\rangle$ using a Krylov subspace method.

\subsubsection*{Rabi-map construction}

For each detuning offset, we compute the observable expectation value using the vec-trace identity,
\begin{equation}
\langle O(t)\rangle
= \mathrm{Tr}[O\rho(t)]
= \mathrm{vec}(O^T)^\dagger\,|\rho(t)\rangle\rangle,
\end{equation}
with $O=\tau_z$ for the spin-conserving transition and 
$O=\sigma_z$ for the spin-flip transition.

Repeating this calculation over a two-dimensional grid of detuning and time produces the Rabi maps presented in Fig.~4 of the main text.
In addition, in our Rabi-oscillation simulations, we took $T_1 = 750~\mathrm{ps}$, the maximum relaxation time obtained from the detuning-energy spectra in Fig.~\ref{FigS4}(a). 
Because the dephasing time must satisfy 2$T_2 \leq T_1$, we used $T_2 = T_1/2$ in our calculation.

As we mentioned in \ref{detuing_spectra}, relaxation time is extracted from the detuning-dependent energy spectrum by identifying spin-conserving and spin-flip transition points where Fano resonances emerge. From the linewidth of these resonances, we determine an effective relaxation timescale. We emphasize that this $T_1$ does not represent the intrinsic coherence limit of graphene, but rather serves as an effective parameter describing environment-induced dissipation, allowing us to investigate the Rabi dynamics within a relevant operational regime.

Furthermore, as shown in Fig.~4 of the Supplementary Materials of Ref.~[2], increasing the junction width $\xi$ leads to more pronounced and eventually saturated resonance peaks near $E \approx \pm 0.04$, accompanied by a clear narrowing of their linewidths. This behavior indicates that in smoother and more realistic potential profiles, the effective dissipation rate is reduced, corresponding to an increase in the relaxation time. While a quantitatively accurate determination of relaxation times would require a more detailed treatment of environmental coupling and noise, which is beyond the scope of this work, our results demonstrate that the proposed device operates within a physically reasonable regime and highlight its potential as a viable spin-qubit platform.

\bibliographystyle{cip-v3-bst-submit}  
\bibliography{references_251030}